\documentclass[12pt,reqno]{amsart}
\usepackage{amsmath,amssymb,amsfonts,amsthm}
\usepackage[mathscr]{eucal}
\usepackage[all]{xy}
\usepackage{hyperref}
\usepackage{setspace}
\allowdisplaybreaks

\author{D.S.~Kaparulin, S.L.~Lyakhovich}

\address{Department of Quantum Field Theory, Tomsk State University, Lenin ave. 36, Tomsk 634050, Russia.}

\allowdisplaybreaks

\title{Unfree gauge symmetry in the BV formalism}

\begin{document}

 \maketitle

\begin{abstract} The BV formalism is proposed for the theories where the gauge symmetry parameters are unfree, being constrained by differential equations.
\end{abstract}

\section{Introduction}
The  Batalin-Vilkovisty (BV) formalism\footnote{Also known as BRST
(Becci-Ruet-Stora-Tutin) field-anti-field formalism} was initially
proposed \cite{BV1}, \cite{BV2}, \cite{BV3}, \cite{BV4} as a tool
for quantizing classical gauge field theories. Later on, the scope
of applications of the formalism has been extended to a large
variety of problems in physics and mathematics ranging from
consistent inclusion of interactions in gauge field theories to the
characteristic classes of various manifolds.

Given the Lagrangian, the BV-BRST embedding of the theory is a
well-known straightforward procedure  \cite{Henneaux:1992ig} if
certain regularity conditions are obeyed by the original field
equations and their gauge symmetry. These regularity conditions are
also generalized for not necessarily Lagrangian field equations to
provide their BV-BRST embedding \cite{Lyakhovich:2004xd},
\cite{Kazinski:2005eb}. We mention two conditions which are assumed
to hold true for the original field theory to admit the usual
BV-BRST embedding: (i) the gauge symmetry parameters are arbitrary
functions of space-time coordinates, i.e. they are not constrained
by any equations; (ii) any on-shell vanishing function of the fields
and their derivatives can be spanned by the left hand sides of the
field equations and their differential consequences. These two
assumptions, being critical for the BV formalism construction, are
violated in some field theory models of a current interest. Examples
are given in Section 6. As one can see from the examples, both the
assumptions are usually violated simultaneously. Once the gauge
transformation parameters are constrained by equations, the gauge
symmetry is named unfree.

In the recent article \cite{Kaparulin:2019quz}, the defining
relations are found of the unfree gauge symmetry algebra. The
algebra of gauge symmetry with unconstrained gauge parameters is
constructed starting from two basic ingredients: the action
functional and gauge symmetry  generators. In the unfree case, two
more basic ingredients are involved: the operators of gauge
parameter constraints and the mass shell completion functions. The
first extra ingredient defines the equations constraining gauge
symmetry parameters. The completion functions constitute the
generating set of the on-shell vanishing quantities such that do not
reduce to the left hand sides of the Lagrangian equations and their
differential consequences. Proceeding from the simplest case of the
general unfree gauge algebra such that there is no off-shell
disclosure, the Faddeev-Popov (FP) quantization recipe is deduced in
the article \cite{Kaparulin:2019quz}. Earlier, in the specific case
of the unimodular gravity, the FP quantization recipe has been
deduced in the paper \cite{Percacci1}, \cite{Percacci2} making use
of some nonlocal manipulations involving splitting the fields into
longitudinal and transverse components and special gauge conditions.
The set of ghosts involved in the general FP recipe in the case of
unfree gauge parameters includes some extra variables comparing to
the case of the unconstrained gauge symmetry. In the examples where
the models with unfree gauge symmetry admit equivalent
reformulations with unconstrained gauge parameters, the modified FP
recipe can be explicitly reduced to the standard one
\cite{Kaparulin:2019quz}.

In this paper, we propose the extension of the BV formalism to a
general Lagrangian theory with unfree gauge symmetry. Our focus is
at the algebraic aspects of the extension, while the subtleties are
left aside concerning the functional aspects.  In the next section,
we describe the algebra of unfree gauge symmetry. Proceeding from
the relaxed regularity assumptions such that admit irreducible
unfree gauge symmetry, we deduce the basic structure relations of
the gauge algebra including the most general off-shell terms. In
Section 3, we propose the BV embedding for the Lagrangian theory
with unfree gauge symmetry. This requires to introduce the set of
ghosts and anti-fields adjusted for the unfree gauge algebra. The
ghost and anti-field set is different from the theory with
unconstrained gauge parameters. Given the set of fields and
anti-fields, and the relaxed regularity conditions, we see that the
classical master equation reproduces the structure relations of
unfree gauge symmetry algebra. In Section 4, we consider the
non-minimal sector and gauge fixing. As one can notice from the
examples, the specifics of the unfree gauge symmetry is that the
independent gauge fixing conditions break the relativistic symmetry,
while the relativistic gauges are inevitably redundant even though
the unfree gauge symmetry is irreducible. We give the non-minimal
sector ghosts both for the independent and redundant gauges. For the
theory without off-shell disclosure, we deduce the FP recipe by
explicitly fixing the anti-fields by the gauge conditions. This is
done both with independent and redundant gauges. In Section 5, we
prove the existence theorem for the unfree gauge algebra. We use the
homological perturbation theory (HPT) method. The key ingredient of
the method is the Koszul-Tate differential. In the context of the
existence theorem for the BV master equation, the Koszul-Tate
differential has been first considered in the works \cite{BV4} and
\cite{VT1982}. The HPT method based on the Koszul-Tate complex was
formulated in the article \cite{Fisch:1989rp} as a tool for BV
embedding of general Lagrangian gauge theories. This HPT procedure
follows the pattern earlier suggested in the work
\cite{Fisch:1989dq} for the
BFV\footnote{Batalin-Fradkin-Vilkovisky}-BRST embedding of the
Hamiltonian systems with reducible first class constraints. For the
basics of the HPT applications to the BV formalism, we refer to the
book \cite{Henneaux:1992ig}. In the case of the unfree gauge
symmetry, the Koszul-Tate resolution of the mass shell differs from
that for the gauge theory with unconstrained gauge parameters. Once
the resolution is identified, the HPT allows one to construct the BV
master action. In Section 6, we review the specific models with
unfree gauge symmetry. After that, we consider one simple model to
exemplify all the stages of the BV construction for the theory with
unfree gauge symmetry. Section 7 includes concluding remarks.

\vspace{0.2 cm}

\noindent{\bf Condensed notations.} In this article except for
Section 6, where the specific field theory models are discussed, we
adopt the DeWitt condensed notation. In this notation, the space of
all the field histories is mimicked by the finite dimensional
manifold $\mathcal{M}$, while the fields $\phi^i$ are treated as the
local coordinates on $\mathcal{M}$. The index $i$ is ``condensed",
i.e. it comprises the space-time argument $x$ of the field, and all
the discrete labels (like tensor, spinor, or flavor indices). The
fields are supposed to obey certain boundary conditions when the
spacial part of $x$ tends to infinity, and the asymptotics are
understood as a part of definition of $\mathcal{M}$. In this article
we imply zero boundary conditions. All the other variables, like
gauge transformation parameters, ghosts, anti-fields, are treated
like if they were the coordinates on the fibers of appropriate
bundles over $\mathcal{M}$. Summation over the condensed index
includes integration over the space-time argument. The matrices with
condensed indices represent differential operators. For example,
$\delta_{ij}$ includes $\delta(x-y)$ and delta symbol of discrete
labels, so the matrix $M_{ij}$ can represent D'Alembertian with
appropriate identification: $i=x, j=y$, $M_{ij}=\Box \delta(x-y)$.
In this notation, the Klein-Gordon equation reads
$(M_{ij}+m^2\delta_{ij})\phi^i=0$. The local functionals, being
integrals of the functions of the fields and their space-time
derivatives $F(\phi)=\int dx\, \mathcal{F}(\phi,
\partial_x\phi, \partial^2_x\phi, \ldots ,
\partial^k_x\phi)$ are mimicked by functions on $\mathcal{M}$,
so the linear space of local functionals is treated like it was just
a special subspace of smooth functions on $\mathcal{M}$:
$F\in\mathcal{C}^\infty(\mathcal{M})$. We name the smooth functions
of the fields and their derivatives $O(\phi,
\partial_x\phi, \partial^2_x\phi, \ldots , \partial^k_x\phi)$  as local functions.
We denote the algebra of local functions as $R(\mathcal{M})$. Any
local function can be viewed as a local functional, so
$R(\mathcal{M})\subset\mathcal{C}^\infty(\mathcal{M})$. Derivatives
with respect to $\phi^i$ are understood as functional derivatives of
any local functional, including any local function. The condensed
notation can be always unambiguously uncondensed. For further
details of the condensed notation, we refer to the books
 \cite{Henneaux:1992ig}, \cite{DeWitt:1965jb}.

\section{Algebra of unfree gauge symmetry}
In this section, we at first address the issue of the Noether
identities and their consequences, in the theories where the on
shell vanishing local functions are not exhausted by the linear
combinations of the left hand sides of the Lagrangian equations. In
the second instance, we demonstrate that the modification of the
Noether identities lead to the unfree gauge symmetry. Proceeding
from the modified identities and unfree gauge symmetry, we deduce
the higher structure relations of the unfree gauge symmetry algebra.

\vspace{0.2cm} Consider the theory of fields $\phi^i$, with the
action $S(\phi)$ being the local functional. The stationary surface
of the action is defined by the Lagrangian equations
\begin{equation}\label{LE}
    \partial_i S(\phi)=0 \, .
\end{equation}
The set of solutions of (\ref{LE}) is usually named the mass shell.
We denote the mass shell $\Sigma$,
\begin{equation}\label{Sigma}
 \Sigma\subset\mathcal{M}\, , \qquad  \Sigma=\{ \phi_0 \in{\mathcal{M}} \, | \, \partial_i S(\phi_0)=0\}
\end{equation}
We use the sign $\approx$ to denote the on shell equality
\begin{equation}\label{approx}
    A(\phi)\approx B(\phi) \quad\Leftrightarrow \quad
    A(\phi_0)=B(\phi_0)\,,\,\quad\forall\phi_o\in\Sigma \,, \quad
    A,B\in R(\mathcal{M}).
\end{equation}
The local function $T(\phi)$ is considered as trivial if it vanishes
on shell.  In the classical field theory with equations of motion
(\ref{LE}), every two local functions are considered equivalent if
they differ by a trivial function,
\begin{equation}\label{equiv}
  A\sim B \quad\Leftrightarrow \quad
    A-B\approx 0\,,\, \quad
    A,B\in R(\mathcal{M}).
\end{equation}
 The trivial local functions constitute an ideal $I(\mathcal{M})\subset R(\mathcal{M})$ in the algebra of local functions. The usual
assumption of the general gauge field theory is that any trivial
local function is spanned by the left hand sides of Lagrangian
equations, i.e. the ideal $I$ is generated by $\partial_i S$,
\begin{equation}\label{T}
    T(\phi)\approx 0 \quad\Leftrightarrow\quad T(\phi)=T^i(\phi)\partial_i
    S(\phi)\, .
\end{equation}
Once the condensed notation is used here, the coefficients
$T^i(\phi)$ can describe the differential operators in space time
argument, with the coefficients being local functions. This can be
said in a slightly different wording. Any relation stating that the
local function vanishes on shell should be the differential
consequence of the Lagrangian equations.

The BV formalism \cite{BV1}-\cite{Henneaux:1992ig} relays on the
assumption (\ref{T}) in several crucial aspects. Let us mention two
of them. First, the restriction (\ref{T}) is included into the set
of sufficient conditions that ensure the existence of solution to
the BV master equation. Second, even if the solution exists for the
master-equation, while (\ref{T}) does not hold true, the ideal $I$
of trivial local functions would not be isomorphic to the BRST-exact
functions of zero ghost number. This would break the usual physical
interpretation of the BRST cohomology groups.

The assumption (\ref{T}) is violated in a number of field theories,
see the examples in Section 6. Below, we elaborate on the general
gauge symmetry algebra with a relaxed assumption (\ref{T}).

If the left hand sides of the Lagrangian equations (\ref{LE}) cannot
span every trivial local function, we assume that the ideal $I$
still admits a finite generating set. The ideal of the on-shell
vanishing functions is supposed to be generated by $\partial_i
S(\phi)$ and by a finite set of the other trivial functions $\tau_a
(\phi)$,
\begin{equation}\label{tau-a}
    \tau_a(\phi)\approx 0 \,, \quad  \tau_a(\phi)\neq K_a^i(\phi)\partial_i
    S(\phi) \, .
\end{equation}
\begin{equation}\label{Completion}
T(\phi)\approx 0\quad \Leftrightarrow\quad T(\phi)=V^i(\phi)
\partial_i S(\phi) + V^a(\phi) \tau_a(\phi) \, .
\end{equation}
Here, $V^i(\phi), V^a(\phi)$ stand for the differential operators in
the space-time argument with the coefficients being local functions
of the fields, as the condensed notations are applied. We name
$\tau_a$, being the non-Lagrangian generating elements of the ideal
of on-shell vanishing local functions, the \emph{completion
functions} of the Lagrangian system (\ref{LE}).

Note that the mass shell $\Sigma$ (\ref{Sigma}) is defined by the
Lagrangian equations (\ref{LE}) while the relations
$\tau_a(\phi)\approx 0 $ do not restrict the solutions of
(\ref{LE}), even though they are not differential consequences of
the equations (\ref{LE}).
 The examples (see Section 6) demonstrate that this
can happen in the field theories which do not reveal any
inconsistency. Also notice that the completion functions
$\tau_a(\phi)$ are defined modulo the left hand sides of the
Lagrangian equations
\begin{equation}\label{tplusds}
    \tau'_a(\phi)\sim\tau_a(\phi)\, , \quad \tau'_a(\phi) =\tau_a(\phi)+\theta_a^i(\phi) \partial_iS(\phi) \, .
\end{equation}

The generating elements of the ideal of on-shell vanishing local
functions can be dependent, i.e. some of their linear combinations
can vanish off shell:
\begin{equation}\label{GI}
\Gamma_\alpha^i(\phi)\partial_iS(\phi)+\Gamma_\alpha^a(\phi)\tau_a(\phi)\equiv
0 \, .
\end{equation}
We consider these relations as the modified Noether identities.

The generators of the identities are considered equivalent
$\Gamma\sim\Gamma'$ if they differ by a trivial generator such that
vanishes on shell:
\begin{eqnarray}
  \Gamma'^i_\alpha(\phi)-\Gamma^i_\alpha(\phi)&=&  E^{ij}_\alpha
(\phi)\partial_iS (\phi)+  E^{ia}_\alpha (\phi)\tau_a(\phi)\, , \quad  E^{ij}_\alpha = -E^{ji}_\alpha;  \label{GG`1}\\
 \Gamma'^a_\alpha(\phi)-\Gamma^a_\alpha(\phi)&=&  E^{ab}_\alpha
(\phi)\tau_b(\phi) -  E^{ia}_\alpha(\phi) \partial_iS (\phi) \, ,
\quad  E^{ab}_\alpha = -E^{ba}_\alpha. \label{GG`2}
\end{eqnarray}
Also notice that the different choice of completion functions
(\ref{tplusds}) results in corresponding transformation of the
modified Noether identity generators:
\begin{equation}\label{GdSchange}
    \tau_a(\phi)\, \mapsto \, \tau'_a(\phi)=\tau_a(\phi)+\theta_a^i(\phi)\partial_iS(\phi) \, , \quad \Gamma^i_\alpha\,\mapsto\, \Gamma'^i_\alpha=\Gamma^i_\alpha+\theta_a^i(\phi)\Gamma^a_\alpha
\end{equation}

We assume that the set of the modified Noether identities (\ref{GI})
is complete, i.e. any identity generator is spanned by the
generators $\Gamma_\alpha$, modulo the equivalence relations
(\ref{GG`1}), (\ref{GG`2}):
\begin{equation}\label{G-complete}
    L^i(\phi)\partial_iS(\phi)+L^a(\phi)\tau_a(\phi)\equiv 0\quad
    \Rightarrow\quad L^i(\phi)\approx
    k^\alpha(\phi)\Gamma^i_\alpha(\phi),\,\,\,L^a(\phi)\approx
    k^\alpha(\phi)\Gamma^a_\alpha(\phi) .
\end{equation}
In this article, we further assume that the modified Noether
identities (\ref{GI}) are not redundant
\begin{equation}\label{GT-irred}
    K^\alpha(\phi)\Gamma_\alpha^i(\phi)\approx
    0\,,\,\,  K^\alpha (\phi)\Gamma_\alpha^a(\phi)\approx
    0 \quad\Leftrightarrow\quad K^\alpha(\phi)\approx 0\, ,
\end{equation}
i.e. there are no identities among the identities. This assumption
can be relaxed. In the case of the theory with unfree gauge
symmetry, the identities for identities could be accounted for, if
they occurred, along the same lines as in the gauge theories with
unconstrained gauge parameters and dependent gauge generators
\cite{BV2}, \cite{BV3}, \cite{Henneaux:1992ig}.

We adopt one more regularity assumption\footnote{This assumption is
obeyed by all the presently known theories with unfree gauge
symmetry. The examples are provided in Section 6. It could be be
relaxed, however this would lead to a more involved set of ghosts
and anti-fields needed for the proper BV-BRST embedding of the
theory. At the moment, this option seems having only the academic
interest.} that all the completion functions (\ref{tau-a}) are
essentially involved in the Noether identities (\ref{GI}), i.e.
\begin{equation}\label{Constr-ired0}
K_a(\phi)\Gamma^a_\alpha(\phi)\approx 0 \Leftrightarrow
K_a(\phi)\approx 0 \, .
\end{equation}
Let us detail the off shell consequences of this condition. Once any
on-shell vanishing local function is a linear combination of
$\partial_iS$ and $\tau_a$, these relations mean that off shell
$K_a\Gamma^a_\alpha$ reads
\begin{equation}\label{Constr-irred1}
K_a\Gamma^a_\alpha\equiv K_\alpha^a\tau_a+ K^i_\alpha\partial_i S\,,
\end{equation}
while $K_a$ is also spanned by the completion functions and the
l.h.s. of Lagrangian equations:
\begin{equation}\label{Constr-irred2}
K_a=L^b_a \tau_b + L^i_a\partial_i S \, .
\end{equation}
Substituting $K_a$ from (\ref{Constr-irred2}) into
(\ref{Constr-irred1}) we get the Noether identity between the
Lagrangian equations and completion functions (\ref{tau-a}). As any
set of Noether identities is spanned by the generators $\Gamma$
(\ref{G-complete}), we arrive at the relations connecting the
coefficients in the right hand sides of the relations
(\ref{Constr-irred1}) and (\ref{Constr-irred2}):
\begin{equation}\label{Constr-irred3}
K_\alpha^a =L^a_b \Gamma^b_\alpha +
M^\beta_\alpha\Gamma^a_\beta+A^{ab}_\alpha\tau_b+
A^{ai}_\alpha\partial_i S \,,\qquad A^{ab}=-A^{ba}\, ;
\end{equation}
\begin{equation}\label{Constr-irred4}
K_\alpha^i =L^i_b \Gamma^b_\alpha +
M^\beta_\alpha\Gamma^i_\beta-A^{ai}_\alpha \tau_a +
A^{ij}_\alpha\partial_j S \, ,\qquad A^{ij}=-A^{ji}\,.
\end{equation}

\vspace{0.2cm} \noindent Now, let us turn to the issue of gauge
symmetry. Consider the infinitesimal transformation of $\mathcal{M}$
\begin{equation}\label{RT}
    \delta_\epsilon\phi^i=R^i_\alpha(\phi)\epsilon^\alpha \, .
\end{equation}
The infinitesimal parameters $\epsilon^\alpha$ are labeled by the
condensed index $\alpha$ that means they are functions of the
space-time argument. The transformation (\ref{RT}) is understood as
the gauge symmetry if it leaves the action invariant,
\begin{equation}\label{RS}
    \delta_\epsilon
    S(\phi)\equiv\epsilon^\alpha R^i_\alpha\partial_iS(\phi)\equiv 0
    \,  .
    \end{equation}
As the mass shell (\ref{Sigma}) is a stationary surface of the
action, $\Sigma$ is automatically invariant under the gauge
transformation. The gauge invariance condition (\ref{RS}) of the
action means that certain linear combinations  identically vanish of
the Lagrangian equations. It is the second Noether theorem. Once the
system does not obey the assumption (\ref{T}), the most general
Noether identity (\ref{GI}) involves both the Lagrangian equations
(\ref{LE}) and completion functions (\ref{tau-a}). Any other Noether
identity would be a linear combination of the ones from the
generating set of identities (\ref{G-complete}). With this regard,
one can identify the most general gauge symmetry generator
$R^i_\alpha$ (\ref{RT}), (\ref{RS}) with the generator of Noether
identity $\Gamma^i_\alpha$ (\ref{GI}). Then, under the
transformation
\begin{equation}\label{GT}
    \delta_\epsilon\phi^i=\epsilon^\alpha\Gamma^i_\alpha(\phi)
\end{equation}
the action transforms as follows, given the identity (\ref{GI}):
\begin{equation}\label{GTS}
  \delta_\epsilon S(\phi)\equiv\epsilon^\alpha\Gamma^i_\alpha(\phi)\partial_iS(\phi) \equiv - \epsilon^\alpha\Gamma^a_\alpha(\phi)\tau_a(\phi)\,.
\end{equation}
As all the operators $\Gamma^a_\alpha$ are assumed
 independent (\ref{GT-irred}), this means the action is invariant
 provided for the gauge parameters are constrained by the equations
\begin{equation}\label{eps-constr}
\epsilon^\alpha\Gamma^a_\alpha(\phi)=0.
\end{equation}
In the other wording,  once the gauge variation (\ref{GT}) is unfree
of the fields, with the gauge parameters obeying the equations
(\ref{eps-constr}), the action functional is invariant under the
gauge transformation,
\begin{equation}\label{S-inv}
     \delta_\epsilon
S(\phi)\equiv 0 \, .
\end{equation}
We see that the modified Noether identities (\ref{GI}) involving
completion functions (\ref{tau-a}) result in the unfree gauge
symmetry of theory.

In the gauge identities (\ref{GI}),  the operators $\Gamma^i_\alpha$
and $\Gamma^a_\alpha$ are involved on an equal footing. However,
they have different roles in the gauge symmetry transformations. The
operator $\Gamma^i_\alpha$ generates the unfree gauge
transformations (\ref{GT}), while $\Gamma^a_\alpha$ defines the
equations (\ref{eps-constr}) that restrict the gauge parameters.
With this regard, we name $\Gamma^i_\alpha$ the generators of unfree
gauge symmetry, while $\Gamma^a_\alpha$ are named the operators of
gauge parameter constraints.

The local function(al) $O(\phi)$ is considered gauge invariant if
the unfree gauge variation vanishes of $O(\phi)$ on shell,
\begin{equation}\label{delta-O}
  \delta_\epsilon O(\phi)=\epsilon^\alpha\Gamma^i_\alpha(\phi)\partial_i O(\phi)\approx 0
\end{equation}
Given the regularity conditions  (\ref{Completion}),
(\ref{G-complete}), (\ref{GT-irred}) imposed on the mass shell
(\ref{Sigma}) and the gauge parameter constraints
(\ref{eps-constr}), this relation can be formulated off-shell, and
without explicit involvement of the unfree parameters:
\begin{equation}\label{Off-sh}
  \delta_\epsilon O(\phi)\approx 0
   \quad\Leftrightarrow\quad\Gamma^i_\alpha\partial_iO(\phi)+V^i_\alpha(\phi)\partial_iS(\phi)
   +V^a_\alpha(\phi)\tau_a(\phi)+ W_a(\phi)\Gamma^a_\alpha(\phi)\equiv 0 \,.
\end{equation}
Once the mass shell is invariant under the unfree gauge
transformations, the ideal of the on-shell vanishing local functions
is also gauge invariant,
 \begin{equation}\label{shell-inv}
    T(\phi)\approx 0 \quad\Rightarrow\quad \delta_\epsilon
    T(\phi)\approx 0 \, .
\end{equation}

The relation (\ref{shell-inv})  applies to any element of the ideal,
including the completion functions,
$\delta_\epsilon\tau_a(\phi)\approx 0$. Making use of (\ref{Off-sh})
we get
\begin{equation}\label{Gtau}
\Gamma^i_\alpha(\phi)\partial_i \tau_a(\phi)= R_{\alpha
a}^i(\phi)\partial_iS(\phi)+R_{\alpha a}^b(\phi)\tau_b(\phi)+
W_{ab}(\phi)\Gamma^b_\alpha (\phi)
 \, .
\end{equation}
The last term in this relation does not necessarily vanish on-shell.
It has some specificity comparing to the unfree gauge transformation
of the general on-shell vanishing function (\ref{Off-sh}),
(\ref{shell-inv}): the structure coefficient $W_{ab}$ is on shell
symmetric,
\begin{equation}\label{W-symm}
    W_{ab}(\phi) - W_{ba}(\phi)\approx 0 \, .
\end{equation}
This property can be deduced as a differential consequence of the
identities (\ref{GI}) with the account for the regularity conditions
(\ref{Constr-ired0}), (\ref{Constr-irred1}), (\ref{Constr-irred2}).
Given the regularity conditions
(\ref{Constr-ired0})-(\ref{Constr-irred4}), the differential
consequences of the identities  (\ref{GI}) also define the
commutators of gauge symmetry generators $\Gamma^i_\alpha$, and the
action of the generators onto the operators of gauge parameter
constraints $\Gamma^a_\alpha$:
\begin{eqnarray}\label{GiGi}
\nonumber
\Gamma^i_\alpha(\phi)\partial_i\Gamma^j_\beta(\phi)-\Gamma^i_\beta(\phi)\partial_i\Gamma^j_\alpha(\phi)&=&
    U_{\alpha\beta}^\gamma(\phi)\Gamma^j_\gamma(\phi)+\\
    E_{\alpha\beta}^{aj}(\phi)\tau_a(\phi) + E_{\alpha\beta}^{ij}(\phi)
    \partial_iS(\phi) &+& R_{\alpha a}^{j}(\phi)\Gamma^a_\beta(\phi) - R_{\beta a}^{j}(\phi)\Gamma^a_\alpha(\phi) \,. \label{GG}
\end{eqnarray}
\begin{eqnarray}\label{GiGa}
\nonumber
\Gamma^i_\alpha(\phi)\partial_i\Gamma_\beta^a(\phi)-\Gamma^i_\beta(\phi)\partial_i\Gamma_\alpha^a(\phi)
&=&
 U_{\alpha\beta}^\gamma(\phi)\Gamma_\gamma^a(\phi) + \\
  R_{\alpha}{}_ b^a(\phi)\Gamma^b_\beta(\phi) - R_{\beta}{ }^a_b(\phi)\Gamma^b_\alpha(\phi)&+&
  E_{\alpha\beta}^{ab}(\phi)\tau_b(\phi) + E_{\alpha\beta}^{ai}(\phi)\partial_iS(\phi) \,,
\end{eqnarray}
where the structure functions $E$ are antisymmetric,
$E_{\alpha\beta}^{ij}=-E_{\alpha\beta}^{ji},
E_{\alpha\beta}^{ab}=-E_{\alpha\beta}^{ba}$. The off-shell relations
(\ref{GiGi}) mean, in particular, that any two unfree gauge
transformations (\ref{GT}), (\ref{eps-constr}) commute on-shell to
another unfree gauge transformation. The off-shell relations
(\ref{GiGa}) ensure that the equations (\ref{eps-constr})
constraining the gauge parameters are on-shell gauge invariant
themselves. This allows one to conclude that the unfree gauge
symmetry transformations define on-shell integrable distribution. It
foliates the mass shell into the gauge orbits, much like the gauge
transformations would do if the gauge parameters were not
constrained. This allows one to define physical observables in the
usual way, as the equivalence classes (\ref{equiv}) of the on-shell
gauge invariant local function(al)s (\ref{delta-O}), (\ref{Off-sh}).
Any two observables are considered equivalent if they coincide on
shell (\ref{equiv}). Let us denote the subalgebra of on-shell gauge
invariant local functions as $G(\mathcal{M})$, i.e.
\begin{equation}\label{G-inv}
A\in G\quad \Leftrightarrow\quad \delta_\epsilon A\in I,
\end{equation}
Then the algebra of physical observables is understood as a quotient
algebra $G/I$.

Let us summarize the most important specifics of the unfree gauge
symmetry algebra. First, the generating set for the ideal of trivial
local functions is not exhausted by the left hand sides of
Lagrangian equations, it also includes the completion functions
(\ref{tau-a}). The Noether identities are modified (\ref{GI}) also
involving completion functions. Second, the gauge transformation
parameters (\ref{GT}) are unfree, being constrained by the equations
(\ref{eps-constr}). Third, the regularity/completeness assumptions
involve the equations of motion, completion functions, gauge
generators and gauge parameter constraint operators
(\ref{Completion}), (\ref{G-complete}), (\ref{GT-irred}). This
specifics has to be accounted by an appropriate modification of the
BV formalism such that can cover the systems with unfree gauge
symmetry.

\vspace{0.2cm}

\textbf{Remark.} Let us finalize the section with a remark on the
possible alternative parametrization of the gauge symmetry such that
have unconstrained gauge parameters. Notice the general solution to
the equations constraining the gauge parameter (\ref{eps-constr})
should involve the arbitrary functions.\footnote{The unfree gauge
transformations should not be confused with so-called semi-local
symmetries, see \cite{Sorokin}, \cite{Beckaert}, \cite{Bandos} and
references therein. In both the cases, the transformation parameters
have to obey the differential equations. The difference is that the
solutions to the equations on the unfree parameters involve the
arbitrary functions of $d$ coordinates in $d$ dimensional space,
while in the semi-local case the arbitrary functions depend on $d-1$
coordinates or less.} We denote these arbitrary functions
$\omega^A$. The condensed index $A$ includes the space-time argument
$x$, so $\omega^A$ are the arbitrary functions of $x$ indeed.
 They can be considered as the
unconstrained gauge symmetry transformation parameters. In this
setting, the solution to the equations (\ref{eps-constr}) read
\begin{equation}\label{GT-red}
\exists \Lambda^\alpha_A(\phi)\,: \quad
\epsilon^\alpha\Gamma_\alpha^a(\phi)\approx
    0\quad\Leftrightarrow\quad \epsilon^\alpha\approx \Lambda^\alpha_A(\phi) \omega^A \,
    ,
\end{equation}
The on-shell equality can be extended off shell,
\begin{equation}\label{GT-red-off}
    \Lambda^\alpha_A\Gamma_\alpha^a=E^{ab}_A\tau_b+ E^{ai}_A\partial_i
    S \, , \quad E^{ab}_A=-E^{ba}_A \, .
\end{equation}
Introduce the new generators of gauge symmetry, being linear
combinations of the original ones modulo on-shell vanishing terms:
\begin{equation}\label{G-red}
    G_A^i= \Lambda^\alpha_A\Gamma^i_\alpha + E^{ai}_A\tau_a \, .
\end{equation}
Apply these generators to the action functional,
\begin{equation}\label{GT-S}
G_A^i\partial_iS \equiv \Lambda^\alpha_A\Gamma^i_\alpha\partial_iS +
E^{ai}_A\tau_a\partial_iS .
\end{equation}
The identities (\ref{GI}) mean that $ \Gamma^i_\alpha\partial_iS
\equiv-\Gamma^a_\alpha\tau_a$. Substituting that to (\ref{GT-S}),
and accounting for (\ref{GT-red-off}),  we see
\begin{equation}\label{GI-G}
G_A^i\partial_iS \equiv 0 \, .
\end{equation}
This means, the generators (\ref{G-red}) define the unconstrained
gauge symmetry of the action. Also, notice that the gauge symmetry
transformations
\begin{equation}\label{gt-omega}
    \delta_\omega\phi^i= G_A^i\omega^a
\end{equation}
can be reducible, the symmetry of symmetry can occur. In the article
\cite{Francia:2013sca}, it was shown that the unconstrained local
parametrization of the gauge symmetry always exists for the linear
field theories. This does not mean that any field theory model
reduces to regular theory with (maybe reducible) unconstrained gauge
symmetry. Even though the gauge symmetry transformations
(\ref{gt-omega}) involve unconstrained parameters, the completion
functions do not disappear from the dynamics. They remain on shell
vanishing, and thereby trivial, while they do not reduce to the
linear combinations of the Lagrangian equations. So, if one tried to
find the BV master action along the usual lines of BV formalism for
the regular reducible gauge theory, and ignoring the completion
functions, this can result in the contradictions. The matter is that
the regularity conditions (\ref{T}) are invalid if the the
completion functions (\ref{tau-a}) admitted in the theory, while the
existence theorem \cite{BV4}, \cite{Henneaux:1992ig} implies the
relations (\ref{T}) to hold true for the Lagrangian equations. So,
the completion functions can obstruct the existence of the solution
to the usual master equation. Furthermore, even if the solution
exists, the completion functions will not correspond to the
BRST-exact quantities, while they are trivial. This would violate
the usual physical interpretation of the BRST cohomology. With this
regard, in the next section we consider the construction of the BV
formalism for the unfree gauge symmetry algebra with a proper
account for the role of completion functions.

\section{Master equation}
Construction of the BV-BRST enbedding for the gauge system begins
with the definition of the ghost and anti-field extension of the
original set of the fields $\phi^i$. Below we provide some reasons
for certain ghost/anti-field extension of $\mathcal{M}$ and
formulate the master equation for the action. After that, we shall
see that the master equation indeed reproduces the structure
relations of the unfree gauge symmetry algebra deduced in the
previous section. Then we shall see that the algebra of the gauge
invariants of the original theory is mapped to the BRST cohohomology
of the BV formalism. In the end of the section, we provide a
reinterpretation of the constructed BV formalism in terms of
``compensator fields". In the next section we consider the gauge
fixing in the BV formalism. A formal justification of the specific
ghost and anti-field set is provided in Section 5, where we identify
the Koszul-Tate differential for the mass shell of the theory with
the unfree gauge symmetry and prove the existence theorem for the
master equation.

To make the appropriate choice of ghosts and anti-fields for the
theory with unfree gauge algebra, we proceed from the analogy with
the BRST embedding of the not necessarily Lagrangian systems
\cite{Lyakhovich:2004xd}, \cite{Kazinski:2005eb}. If the theory is
defined just by the equations of motion (not necessarily
Lagrangian), every equation is assigned with the anti-field whose
ghost number is $-n-1$, where $n$ is the ghost number of the
equation. Every generator of gauge identity is also assigned with
the anti-field whose ghost number is $-n-2$, where $n$ is the ghost
number of the equations involved in the identity. Every generator of
the gauge symmetry is assigned with the ghost whose ghost number is
$k+1$, where $k$ is the ghost number of the gauge parameter. Notice
that for the non-Lagrangian systems, the gauge symmetries are not
necessarily paired with the gauge identities, so the corresponding
generators can be different, while in the Lagrangian case the same
operator generates both gauge symmetry and gauge identity. In the
Lagrangian systems with unfree gauge symmetry, the non-Lagrangian
pattern works well for introducing ghosts and anti-fields after two
adjustments. The first is that the anti-fields are to be assigned to
every element of the generating set for the ideal of on-shell
vanishing local functions. This means, the anti-fields are
introduced both for Lagrangian equations (\ref{LE}) and completion
functions (\ref{tau-a}). We denote these anti-fields $\phi^*_i$ and
$\xi^*_a$, respectively. The second is that the ghost $C^\alpha$ is
assigned to every gauge symmetry generator $\Gamma_\alpha^i$
(\ref{GT}) even though the gauge parameters are unfree
(\ref{eps-constr}). The constraints on the gauge parameters are
accounted for by imposing the same constraints on the ghosts:
\begin{equation}\label{gh-constr}
\Gamma^a_\alpha(\phi)C^\alpha=0 \, .
\end{equation}
These equations are not involved in any gauge identity because of
(\ref{Constr-ired0}). We treat the equations for ghosts
(\ref{gh-constr}) on an equal footing with the other generating
elements of $I$. This means, the anti-field has to be introduced for
every ghost constraint (\ref{gh-constr}). As the equations
(\ref{gh-constr}) are of the ghost number  $1$, the anti-fields
should be assigned with the ghost number zero.  We denote these
anti-fields $\xi^a$. It does not mean the mere extension to the set
of original fields $\phi^i$, because the BV formalism also implies
one more grading: the resolution degree (also referred to as the
anti-ghost number in some literature, e.g. in
\cite{Henneaux:1992ig}). The anti-field $\xi^a$ has the resolution
degree $1$, unlike $\phi^i$. And finally, the anti-fields
$C_\alpha^*$ with ghost number $-2$ are assigned to the modified
Noether identities (\ref{GI}) alike the Lagrangian theory with
unconstrained gauge parameters, even though the identities involve
completion functions, unlike the usual case.

Let us introduce the notations for the gradings. The Grassmann
parity is denoted $\varepsilon$, the ghost number is $\text{gh}$,
and the resolution degree is $ \text{deg}$. The assigned gradings
are arranged in Table 1.

\vspace{0.2cm}

\begin{center}
\begin{tabular}{|c|c|c|c|c|c|c|}
  \hline
  grading$\backslash$variable & $\phantom{0}\phi^i$ \phantom{0}& \phantom{0} $\xi^a$ \phantom{0}& \phantom{0} $C^\alpha$ \phantom{0} &
  \phantom{0} $\phi^*_i$ \phantom{0} & \phantom{0}$\xi^*_a$ \phantom{0}& \phantom{0} $C^*_\alpha$ \phantom{0} \\
  \hline
  $\varepsilon $ & 0 & 0 & 1 &  1  &  1&  0\\
  $\text{gh}$        & 0 & 0 & 1 & -1  & -1& -2\\
%$\text{pgh}$       & 0 & $\textbf{1}$ & 1 &  0  &  0&  0 \\
  $\text{deg}$       & 0 & $\textbf{1}$ & 0 &  1  &  1&  2 \\
  \hline
\end{tabular}
\end{center}
\begin{center}
Table 1.
\end{center}
It is also convenient to use the collective notation
\begin{equation}\label{varphi}
    \varphi^I=(\phi^i,\xi^a,C^\alpha)\,,\qquad
    \varphi^\ast_I=(\phi^\ast_i,\xi^\ast_a,C^\ast_\alpha)\,.
\end{equation}
The ghost number grading of the variables is explained above. Now,
we explain the reasons for assigning such resolution degrees to
these variables. All the anti-fields paired to the generating
elements of the ideal $I$ are assigned with the resolution degree
$1$. It is the same principle as in the gauge theory with
unconstrained gauge parameters, with one adjustment: the anti-fields
are introduced not only for the original Lagrangian equations
(\ref{LE}) -- $\phi^\ast_i$, but also to the completion functions
(\ref{tau-a}) -- $\xi^\ast_a$, and to the equations constraining the
ghosts (\ref{gh-constr}) -- $\xi^a$. The anti-fields are also
introduced being paired with the Noether identities (\ref{GI}) --
$C^\ast_\alpha$. These are assigned with the resolution degree $2$,
much alike the theory with unconstrained gauge parameters.

Once the ghosts and anti-fields are introduced, this means we
extended the manifold of original fields $\mathcal{M}$ to a
$Z$-graded manifold $\bar{\mathcal{M}}$. The fields
$\varphi,\varphi^\ast$ are considered as coordinates on
$\bar{\mathcal{M}}$. The original equations of motion (\ref{LE}) and
the constraints imposed on the ghosts (\ref{gh-constr}) define the
extended mass shell $\bar{\Sigma}\subset\bar{\mathcal{M}}$,
\begin{equation}\label{barSigma}
    \bar{\Sigma}=\{\, (\varphi_0,\varphi^*_0)\in\bar{\mathcal{M}}\, |\, \partial_iS(\phi_0)=0,\, C^\alpha_0\Gamma^a_{\alpha}(\phi_0)=0
    \}\, .
\end{equation}
The algebra of local functions on $\bar{\mathcal{M}}$ is denoted as
$\bar{R}$. It includes the ideal of the on-shell vanishing local
functions, $\bar{I}$. The generating set for the ideal consists of
the left hand sides of the Lagrangian equations (\ref{LE}),
completion functions (\ref{tau-a}), and constraints on the ghosts
(\ref{gh-constr}),
\begin{equation}\label{barI}
    A\in\bar{I}\quad\Leftrightarrow\quad
    A=A^i(\varphi,\varphi^\ast)\partial_iS+A^a(\varphi,\varphi^\ast)\tau_a+A_a(\varphi,\varphi^\ast)C^\alpha\Gamma^a_\alpha
    \, .
\end{equation}

The classical theory is Lagrangian because the original mass shell
is defined as a stationary surface of the action, even though the
ideal $\bar{I}$ is not spanned by the left hand sides of Lagrangian
equations. So, we are going to construct the BRST formalism in the
Lagrangian BV form, representing the BRST differential as the
anti-bracket with the master action $S(\varphi,\varphi^\ast)$.

Then, the first question is the definition of the antibracket for
the local function(al)s $A(\varphi,\varphi^\ast)$.  The set of
fields (\ref{varphi}) involves a pair of the anti-fields $\xi^a,
\xi^\ast_a$ that does not have a counterpart in the theories without
constraints on the gauge parameters, while all the other
(anti-)fields have. As these variables are dual to each other, we
find it natural to consider them as conjugate with respect to the
anti-bracket. In the BV formalism, where the ghosts are
unconstrained, once the variable has the positive resolution degree,
being anti-field, the conjugate field always has zero degree. Both
$\xi$ and $\xi^\ast$ are the anti-fields carrying positive degree,
while they are conjugate to each other. As we shall see, this does
not lead to any contradiction, though it may seem unusual. The other
variables are naturally split into the conjugate pairs alike their
counterparts in the BV formalism of the gauge theory without
constraints onto parameters. All these reasons lead us to adopt the
canonical anti-bracket with $\varphi^I$ being conjugate to
$\varphi^\ast_I$:
\begin{equation}\label{anti-br}
    (A,B)=\frac{\partial^RA}{\partial
    \varphi^I}\frac{\partial^LB}{\partial\varphi^\ast_I}-\frac{\partial^RA}{\partial
    \varphi^\ast_I}\frac{\partial^LB}{\partial\varphi^I}\,,\qquad
    \varphi^I=(\phi^i,\xi^a,C^\alpha)\,,\qquad
    \varphi^\ast_I=(\phi^\ast_i,\xi^\ast_a,C^\ast_\alpha)\,.
\end{equation}
The antibracket is Grassmann odd and it shifts the ghost number by
one:
\begin{equation}\label{ab-grad}
    \text{gh}((A,B))= \text{gh}(A)+\text{gh}(B)+1\,, \qquad    \varepsilon((A,B))=
    \varepsilon(A)+\varepsilon(B)+1\, .
\end{equation}
The bracket is inhomogeneous with respect to the resolution degree.

The master action $S(\varphi,\varphi^\ast)$ is defined as an
expansion with respect to the resolution degree
\begin{equation}\label{SBV-exp}
    S=\sum_{k=0}S_k\,, \qquad \text{gh}\,(S_k)\,=\,\varepsilon\,(S_k)\,=\,0\, , \qquad
    \text{deg}\,(S_k)\,=\,k\, ,
\end{equation}
i.e. it is the graded expansion in the anti-fields
$\xi^a,\,\xi^\ast_a, \, \phi^\ast_i, \, C^\ast_\alpha$. The initial
term in this expansion is the original action
\begin{equation}\label{S0}
S_0=S(\phi)\, .
\end{equation}
Given the grading restrictions (\ref{SBV-exp}), the most general
first and second resolution degree terms read
\begin{equation}\label{S1}
    S_1=\tau_a\xi^a+(\phi^*_i\Gamma^i_\alpha + \xi^*_a \Gamma^a_\alpha
    )C^\alpha\,,\phantom{\frac12}
\end{equation}
\begin{equation}\label{S2}
    S_2=\frac12(C^*_\gamma U^\gamma_{\alpha\beta} +
    \phi^*_j\phi^*_iE^{ij}_{\alpha\beta}
    +2\xi^*_a\phi^*_iE^{ia}_{\alpha\beta} + \xi^*_b\xi^*_aE^{ab}_{\alpha\beta} )C^\alpha C^\beta  -\xi^b( \phi^*_i R_ b^i{}_{\alpha} +\xi^*_a R_
    b^a{}_{\alpha})C^\alpha-
    \frac12\xi^b\xi^aW_{ab},
\end{equation}
where all the expansion coefficients $\tau, \Gamma, U,$ etc., can
depend on the original fields $\phi$. The notations for the
coefficients coincide with the corresponding structure functions in
the relations of the unfree gauge symmetry algebra deduced in the
previous section. It is not an abuse of notation. The structures
with identical notation coincide indeed, as we shall see soon.

The master action (\ref{SBV-exp}) is defined by the BV master
equation
\begin{equation}\label{SS}
    (S,S)=0\,.
\end{equation}
The solution to the master equation can be iteratively sought for by
expanding the left hand side with respect to the resolution degree.
 The first two orders of the expansion of the equation involve
 only the first three orders of the master action, i.e. (\ref{S0}),
(\ref{S1}),(\ref{S2}). Explicitly, this reads
\begin{equation}\label{SS0}
    (S,S)_0=2(\Gamma^a_\alpha\partial_i
    S+\Gamma^a_\alpha\tau_a)C^\alpha=0\,,
\end{equation}
\begin{eqnarray}
\nonumber
  (S,S)_1 &=& 2\xi^a(\Gamma^i_\alpha\partial_i\tau_a-R^i_{\alpha
    a}\partial_i S-R^b_{\alpha
    a}\tau_b-W_{ab}\Gamma^b_\alpha)C^\alpha- \\
\nonumber   &- & C^\alpha C^\beta\big(\phi^\ast_i(
\Gamma^j_\alpha\partial_j \Gamma_\beta^i-\Gamma^j_\beta\partial_j
\Gamma_\alpha^i-U^\gamma_{\alpha\beta}\Gamma^i_\gamma-R^i_{\alpha
    a}\Gamma^a_\beta+R^i_{\beta a}\Gamma^a_\alpha-E^{ji}_{\alpha\beta}\partial_j
    S-E^{ia}_{\alpha\beta}\tau_a)- \\
  &-&\xi^\ast_a (\Gamma^j_\alpha\partial_j \Gamma_\beta^a-\Gamma^j_\beta\partial_j \Gamma_\alpha^a-U^\gamma_{\alpha\beta}\Gamma^a_\gamma-R^a_{\alpha
    b}\Gamma^b_\beta+R^a_{\beta b}\Gamma^b_\alpha+E^{ja}_{\alpha\beta}\partial_j
    S-E^{ab}_{\alpha\beta}\tau_b)\big)=0\,. \label{SS1}
\end{eqnarray}
The coefficients at all the independent monomials of the ghosts and
anti-fields should be set to zero separately. In this way, we see
that the zero order of the master equation expansion (\ref{SS0}) is
equivalent to the modified Noether identities (\ref{GI}) of unfree
gauge symmetry algebra. The first order expansion of the master
equation (\ref{SS1}) is equivalent to the structure relations
(\ref{Gtau}), (\ref{GiGi}), (\ref{GiGa}) of the unfree gauge
symmetry algebra. One can see that the set of the ghosts and
anti-fields of Table 1, and the boundary conditions (\ref{S0}),
(\ref{S1}) for the master action,  lead to the solution of the
master equation (\ref{SS}) such that indeed corresponds to the
theory with unfree gauge symmetry algebra. The solution to the
master equation (\ref{SS}) exists in all the orders with respect to
the resolution degree as we shall see in Section 5.

Once the master action is constructed, it defines the BRST
differential
\begin{equation}\label{BRST-diff}
    sA=(A,S)\, .
\end{equation}
$s$ squares to zero because of the master equation (\ref{SS}) and
Jacobi identity for the anti-bracket. It is Grassmann odd vector
field of the ghost number 1,
\begin{equation}\label{s^2}
    s^2=0\,, \qquad \text{gh}(s)=1,\quad \varepsilon(s)=1\, .
\end{equation}
The BRST differential can be expanded with respect to the resolution
degree
\begin{equation}\label{s-exp}
    s=\delta+\gamma+\stackrel{_{(1)}}{s}\ldots\,,\qquad
    \text{deg}\,\delta=-1\,,\qquad \text{deg}\,\gamma=0\,,\qquad
    \text{deg}\,\stackrel{_{(1)}}{s}=1\,.
\end{equation}
As a consequence of (\ref{s^2}), for the lower order terms of the
expansion (\ref{s-exp}) we have
\begin{equation}\label{delta-gamma}
    s^2=0\,\qquad \Rightarrow \qquad \delta^2=0\,,\qquad
    \delta\gamma+\gamma\delta=0\,,\qquad
    \gamma^2+(\delta\stackrel{_{(1)}}{s}+\stackrel{_{(1)}}{s}\delta)=0\,,\qquad
    \ldots
\end{equation}
As one can see, $\delta$, being lowest resolution degree component
of the BRST differential $s$, squares to zero, so it is a
differential in itself. Explicitly, $\delta$ reads
\begin{equation} \label{KT-p}
 \delta A=-\frac{\partial^R A}{\partial \phi^*_i}\partial_iS - \frac{\partial^R A}{\partial
    \xi^*_a}\tau_a+\frac{\partial^R A}{\partial
    C^*_\alpha}\left( \phi^*_i\Gamma^i_\alpha + \xi^*_a\Gamma^a_\alpha\right)+\frac{\partial^R A}{\partial \xi^a}\Gamma^a_\alpha C^\alpha\, .
\end{equation}
It squares to zero, and this is equivalent to the modified Noether
identities (\ref{GI}),
\begin{equation}\label{delta^2}
     \delta^2A\displaystyle=-\frac{\partial^R A}{\partial
    C^*_\alpha}\left(\Gamma^j_\alpha\partial_jS+\Gamma^a_\alpha\tau_a
    \right)=0\,.
\end{equation}
The differential $\delta$ can be understood  as a Koszul-Tate
resolution  for the ideal $\bar{I}$ because of the two reasons. The
first, any on-shell vanishing function of zero resolution degree is
obviously $\delta$-exact
\begin{equation}\label{KT-shell}
\text{ deg} A= 0,\,\, A\in\bar{I} \quad \Leftrightarrow\quad A\equiv
A^i(C,\phi)\partial_iS+A^a(C,\phi)\tau_a+ A_a(\phi,C)\Gamma^a_\alpha
C^\alpha\equiv\delta B\, ,
\end{equation}
where $B=A^a\xi^\ast_a+A_a\xi^a+A^i\phi^\ast_i$. The second,
$\delta$ is acyclic in the strictly positive resolution degree, i.e.
the cohomology of $\delta$ is exhausted by zero resolution degree.
This issue is addressed in Section 5. These two facts allow one to
iteratively find all the higher orders of the master action by the
HPT method.

Zero resolution degree contribution to the BRST differential
(\ref{s-exp}) explicitly reads
\begin{equation}\label{Long}\begin{array}{rl}\displaystyle
    \gamma A&\displaystyle=\frac{\partial^R A}{\partial
    \phi^i} C^\alpha\Gamma^i_\alpha + \frac{1}{2}\frac{\partial^R A}{\partial
    C^\gamma}  U^\gamma_{\alpha\beta}C^\alpha C^\beta +\frac{\partial^R A}{\partial
    \xi^a}((\phi^\ast_iE^{ia}_{\alpha\beta}-\xi^\ast_bE^{ab}_{\alpha\beta})C^\alpha C^\beta-\xi_b R^b_{\alpha a}C^\alpha)-\\[4mm]&\displaystyle
    -\frac{\partial^R A}{\partial
    \phi^*_i}\big(\xi^a\partial_i\tau_a+(\phi^\ast_j\partial_i\Gamma^j_\alpha+\xi^\ast_a\partial_i\Gamma^a_\alpha)C^\alpha\big)+
    \frac{\partial^R A}{\partial\xi^\ast_a}\big((\phi^\ast_iR^i_{\alpha b}+\xi^\ast_aR^a_{\alpha b})C^\alpha+\xi^bW_{ab}\big)
    -\\[4mm]&\displaystyle -\frac{\partial^R A}{\partial C^\ast_{\alpha}}
    \big((U^\gamma_{\alpha\beta}C^\ast_\gamma+\phi^\ast_i\phi^\ast_jE^{ij}_{\alpha\beta}+
    2\phi^\ast_i\xi^\ast_aE^{ia}_{\alpha\beta}+\xi^\ast_a\xi^\ast_b
    E^{ab}_{\alpha\beta})C^\beta+\xi^b(\phi^\ast_iR^i_{\alpha b}+\xi^\ast_aR^a_{\alpha b})\big)\,.
\end{array}\end{equation}
It can be considered as a modification of the longitudinal
differential of the usual gauge theory where the constraints
(\ref{gh-constr}) are not imposed on the ghosts.  Because of
(\ref{s^2}), (\ref{s-exp}) $\gamma$ should anti-commute with
$\delta$. Explicitly, this reads
\begin{equation}\label{delta-gamma2}\begin{array}{l}\displaystyle
    (\delta\gamma+\gamma\delta)A\displaystyle=-\frac{\partial^R
    A}{\partial \phi^\ast_i}\partial_i(\Gamma^i_\alpha\partial_i
    S+\Gamma^a_\alpha\tau_a)C^\alpha-\\[5mm]\displaystyle\qquad-
    \frac{\partial^R A}{\partial
    \xi^\ast_a}(\Gamma^i_\alpha\partial_i\tau_a-R^i_{\alpha
    a}\partial_i S-R^b_{\alpha
    a}\tau_b-W_{ab}\Gamma^b_\alpha)C^\alpha+\\[5mm]\displaystyle\qquad+\frac12\frac{\partial^R
    A}{\partial \xi^a}\big(\Gamma_\alpha^i\partial_i
    \Gamma^a_\beta-\Gamma_\beta^i\partial_i
    \Gamma^a_\alpha-U^\gamma_{\alpha\beta}\Gamma^a_\gamma-R^a_{\alpha b}\Gamma^b_\beta+
    R^a_{\beta b}\Gamma^b_\alpha-E^{ia}_{\alpha\beta}\partial_i S-E^{ab}_{\alpha\beta}\tau_b\big)C^\alpha
    C^\beta-\\[5mm]\displaystyle\qquad-\frac{\partial^R
    A}{\partial C^\ast_\alpha}\bigg[\xi^a(\Gamma^i_\alpha\partial_i
    \tau_a-R^i_{\alpha a}\partial_i S-R^b_{\alpha
    a}\tau_a-W_{ab}\Gamma^b_\alpha)C^\alpha+
    \\[5mm]\displaystyle\qquad+
    \phi^\ast_i(\Gamma^j_\alpha\partial_j
    \Gamma^i_\beta-\Gamma^j_\beta\partial_j\Gamma^i_\alpha-U^\gamma_{\alpha\beta}\Gamma^i_\gamma-R^i_{\alpha b}\Gamma^b_\beta+
    R^i_{\beta
    b}\Gamma^b_\beta-E^{ij}_{\alpha\beta}\partial_j
    S+E^{ia}_{\alpha\beta}\tau_a)+\\[5mm]\displaystyle\qquad+
    \xi^\ast_a(\Gamma^j_\alpha\partial_j
    \Gamma^a_\beta-\Gamma^j_\beta\partial_j\Gamma^a_\alpha-U^\gamma_{\alpha\beta}\Gamma^a_\gamma -R^a_{\alpha b}\Gamma^b_\alpha+
    R^a_{\beta
    b}\Gamma^b_\alpha+E^{ia}_{\alpha\beta}\tau_a-E^{ab}_{\alpha\beta}\partial_i S)\bigg]=0\,,
\end{array}
\end{equation}
The anti-commutator vanishes because of the modified Noether
identities (\ref{GI}), and structure relations of the unfree gauge
symmetry algebra(\ref{Gtau}), (\ref{GiGi}), (\ref{GiGa}).
%\begin{equation}
%\begin{array}{l}\displaystyle
%    \gamma^2A=-\frac{\partial^R
%    A}{\partial\phi^i}(\Gamma^j_\alpha\partial_j\Gamma^i_\beta-U^\gamma_{\alpha\beta}\Gamma^i_\gamma+\ldots)C^\alpha C^\beta+\frac{\partial^R
%    A}{\partial
%    C^\alpha}(\Gamma^i_\beta \partial_i U^\alpha _{\gamma\delta}+U^\alpha_{\beta\varepsilon}U^\varepsilon_{\gamma\delta}+\ldots)C^\alpha
%    C^\beta C^\gamma+\ldots
%\end{array}\end{equation}

Once the BRST differential (\ref{BRST-diff}) has been constructed
such that properly accounts for the completion functions
(\ref{tau-a}) and constraints on the gauge parameters
(\ref{eps-constr}), one can expect that the physical observables of
the theory are defined in the usual way, as the BRST cohomomology
classes of the local function(al)s on $\bar{\mathcal{M}}$ of the
ghost number zero
\begin{equation}\label{H0s}
    H^0(s)=Ker(s)/Im(s)=\{ A \in \bar{R}(\mathcal{\bar{M}}) \,| \, sA=0;\, \text{gh}
    A=0; \, A\sim A',  A-A'=sB \}.
\end{equation}
In the original theory, the algebra physical observables is
understood as the quotient of the subalgebra of on-shell gauge
invariant local function(al)s (\ref{G-inv}) with respect to the
ideal of the on-shell vanishing local function(al)s. This is alike
the theories with unconstrained gauge parameters. The specifics of
the case of unfree gauge symmetry is that the gauge invariance
relations (\ref{Off-sh}) involve the completion functions and gauge
parameter constraint operators. These structures do not have any
analogue in the theories with unconstrained gauge parameters. So,
the question is to verify the isomorphism between the set of
nontrivial gauge invariants of the original theory $G/I$ and BRST
cohomology $H^0(s)$. Cohomological proof of the isomorphism is
provided in Section 5. Here, we just demonstrate the explicit
iterative construction of the BRST invariant corresponding to  gauge
invariant up to the first order with respect to the resolution
degree. Consider the expansion
\begin{equation}\label{O-brst}
    \mathcal{O}(\varphi,\varphi^\ast)=\sum_{k=0}\stackrel{(k)}{\mathcal{O}}\,
    ,\qquad \text{deg}\stackrel{(k)}{\mathcal{O}}=k, \quad
    \text{gh}\stackrel{(k)}{\mathcal{O}}=0\,,
\end{equation}
where $\stackrel{(0)}{\mathcal{O}}=O(\phi)$ is the gauge invariant
of the original theory (\ref{Off-sh}). Given the expansions
(\ref{s-exp}) and (\ref{O-brst}),  the BRST invariance requirement
connects the order $k+1$ with the lower orders of the invariant:
\begin{equation}\label{sO-exp}
    s \mathcal{O}=0 \quad\Leftrightarrow\quad \delta
    \stackrel{(k+1)}{\mathcal{O}}+\gamma\stackrel{(k)}{\mathcal{O}}
    =-\sum_{m=1}^{k}\,
    \stackrel{(m)}{s}\,\, \stackrel{(k-m)}{\mathcal{O}}\, \,,\quad
    k=0,1,2,\ldots
\end{equation}
Consider the most general expression for
$\mathcal{O}(\varphi,\varphi^\ast)$ up to the first order in the
resolution degree
\begin{equation}\label{O1}
    \mathcal{O}(\varphi,\varphi^\ast) =O(\phi)+ \stackrel{(1)}{ \mathcal{O}}\, +\, \ldots = O(\phi)+ C^\alpha
\left(V^i_\alpha\phi^*_i+V^a_\alpha\xi^*_a\right)+ W_a\xi^a +\ldots
\, .
\end{equation}
Substitution of the ansatz (\ref{O1}) into the expansion
(\ref{sO-exp}) leads to the following expression in the leading
order
\begin{equation}\label{gammaO}
    \delta \stackrel{(1)}{\mathcal{O}} + \gamma \stackrel{(0)}{
    \mathcal{O}}=
    C^\alpha\left(V^i_\alpha\partial_iS+V^a_\alpha\tau_a+W_a\Gamma^a_\alpha
    +\Gamma^i_\alpha\partial_i O\right)=0 \, .
\end{equation}
This relation is obeyed by virtue of the gauge invariance condition
for the local function $O(\phi)$ (\ref{Off-sh}) upon identification
of the expansion coefficients $V^i_\alpha,V^a_\alpha, W_a,
\Gamma^a_\alpha, \Gamma^i_\alpha$ for $\stackrel{(1)}{\mathcal{O}}$
in the relation (\ref{O1}) with the corresponding structure
functions in the relation (\ref{Off-sh}).

So, any gauge invariant local function(al) $O(\phi)$ can be extended
to the BRST invariant function(al) at least up to the first order
with respect to the resolution degree. The extension has a natural
ambiguity of the $\delta$-exact terms. This precisely corresponds to
the equivalence relations (\ref{equiv}) for the local function
(\ref{equiv}). So, one can see that the algebra of physical
observables of original theory $G/I$ (\ref{G-inv}) is indeed
isomorphic to the BRST cohomology $H^0(s)$ of zero ghost number  to
the first order in resolution degree. The higher orders exist, and
they all can be iteratively constructed by the HPT method. This is
explained in Section 5.  Let us make one more remark concerning the
solution to master equation. The variable $\xi^a$ has zero ghost
number, and it is Grassmann even if the gauge parameters are even
while it has the resolution degree one. This means, the expansion
can be infinite in $\xi^a$. It might seem not a plausible iterative
procedure for solving the master equation once it does not terminate
in the final number of iterations. In fact, one can avoid explicitly
finding all the orders in $\xi^a$. As we shall see in the next
section, $\xi^a$ can be always fixed by an appropriate choice of
gauge conditions, so only the first orders in $\xi^a$ can matter.

\vspace{0.3 cm}

\textbf{Remark.} Let us finalize the section by the remark on
possible re-interpretation of the constructed BV-BRST formalism for
the original theory with unfree gauge symmetry algebra. Let us
consider the solution $S(\varphi,\varphi^\ast)$ of BV master
equation (\ref{SS})
 for the theory with unfree gauge symmetry
algebra once it has been already constructed with the boundary
conditions (\ref{S0}), (\ref{S1}). If we formally changed the
resolution degree of $\xi^a$ from one to zero, this solution would
remain a solution to the equation (\ref{SS}), because the definition
of the anti-bracket (\ref{anti-br}) does not involve the resolution
degree.
 Then, with respect to the modified resolution degree, the first
two orders read of the master action
\begin{equation}\label{Sxi}
    S(\varphi,\varphi^\ast)=S'(\phi,\xi)+C^\alpha
    R_\alpha^i(\phi,\xi)\phi^\ast_i+
    R_\alpha^a(\phi,\xi)\xi^\ast_a+ \ldots \, .
\end{equation}
This can be considered just as the master action for the theory of
the fields $\phi,\xi$ with the action $S'(\phi,\xi)$ and gauge
symmetry generators $R_\alpha$. As a consequence of the master
equation (\ref{SS}) obeyed by  $S(\varphi,\varphi^\ast)$, the action
$S'(\phi,\xi)$ (\ref{Sxi}) is invariant under the gauge
transformations
\begin{equation}\label{gtxi}
    \delta_\epsilon\phi^i=\epsilon^\alpha R_\alpha^i(\phi,\xi)\,
    ,\quad \delta_\epsilon\xi^a=\epsilon^\alpha
    R_\alpha^a(\phi,\xi)\, ,
\end{equation}
\begin{equation}\label{GS'}
\delta_\epsilon S'\equiv 0\,\quad \forall \epsilon^\alpha\, .
\end{equation}
Notice that the gauge parameters of the transformations (\ref{gtxi})
are not constrained by any equations, it is the usual gauge
symmetry. With this regard, the inclusion of the anti-fields $\xi^a$
to the equations constraining the ghosts (\ref{gh-constr}) can be
re-interpreted as inclusion of the compensator fields.  By extending
the set of original fields with the compensators, the original
theory with the action $S(\phi)$, modified Noether identities
(\ref{GI}), and the unfree gauge symmetry (\ref{GT}),
(\ref{eps-constr}) is converted into the theory of the original
fields $\phi$ and compensators $\xi$ with the unconstrained gauge
symmetry parameters (\ref{gtxi}). It is instructive to expand the
action $S'(\phi,\xi)$ and the gauge generators $R_\alpha(\phi,\xi)$
(\ref{gtxi})  in the power series with respect to the compensator
fields $\xi^a$. The lower orders can be seen of the expansion from
(\ref{S0}), (\ref{S1}), (\ref{S2}):
\begin{equation}\label{S-xiexp}
S'(\phi,\xi)= S(\phi)+\tau_a(\phi)\xi^a + \frac12
W_{ab}(\phi)\xi^a\xi^b +\ldots \,,
\end{equation}
\begin{equation}\label{Rxi}
R^i_\alpha (\phi,\xi)= \Gamma^i_\alpha (\phi) + R^i_{\alpha a}
(\phi) \xi^a +\ldots, \qquad R^a_\alpha (\phi,\xi)= \Gamma^a_\alpha
(\phi) + R^a_{\alpha b}\xi^b +\ldots \, .
\end{equation}
All the expansion coefficients for the action and the gauge
generators are defined by the unfree gauge symmetry algebra
structures described in Section 2. In particular, $\tau_a$ are the
completion functions (\ref{tau-a}); $W_{ab}$ are the structure
coefficients in the relation (\ref{Gtau}); $\Gamma^i_\alpha,
\Gamma^a_\alpha$ are defined by the modified Noether identities
transformations (\ref{GI}) and play the roles of generators of
unfree gauge symmetry transformations (\ref{GT}) and gauge parameter
constraint operators (\ref{eps-constr}); the coefficients $R_{\alpha
a}^i, R^a_{\alpha b}$ are the structure functions in the commutation
relations of the unfree gauge algebra (\ref{GiGi}), (\ref{GiGa}).
So, by introduction of the compensator fields $\xi^a$, the
completion functions and the related structure functions are
absorbed by the action $S'$, while the gauge parameter constraint
operators and related structures are absorbed by the gauge
generators of the extended theory.  All the higher orders of the
expansion (\ref{S-xiexp}), (\ref{Rxi}) exist, because the master
equation has a solution.

The conversion of the theory with the unfree gauge symmetry
parameters into an equivalent theory with extended set of fields,
and without constraints on the gauge parameters is analogous to some
extent to the conversion of the second class constraints into the
first class ones \cite{Batalin:1991jm}, \cite{Batalin:2005df}, or to
inclusion of the St\"uckelberg fields. The difference is that the
second class constraints are absorbed by the first class ones, while
in the theories with unfree gauge parameters, the operators of gauge
parameter constraint operators are absorbed by the gauge generators,
and the completion functions are absorbed by the action. One more
common feature is that the HPT in both cases involves the conjugate
variables such that have non-zero resolution degree, though in the
conversion of Hamiltonian constraints the bracket is even, while in
the unfree gauge theory it is odd.

From the viewpoint of the re-interpretation of the anti-fields
$\xi^a$ as compensators, it is clear that the gauge conditions can
be imposed such that fix $\xi^a$, and even force them to vanish.
With these gauge imposed, it becomes unnecessary to explicitly find
all the orders in $\xi$. It becomes sufficient to explicitly know
the first order in $\xi$, once the higher orders can be excluded
from the gauge fixed BRST invariant action by an appropriate gauge
condition.

\section{Gauge fixing}
In this section, we briefly consider the gauge fixing procedure in
the BV formalism of the theories with unfree gauge symmetry algebra.

The field-anti-field manifold $\mathcal{\bar{M}}$ (see the Table 1)
has a structure of the odd cotangent bundle, and it is equipped with
the canonical anti-bracket (\ref{anti-br}). From this viewpoint, it
is the usual setup of the BV formalism. With this regards, the gauge
fixing should mean the choice of the Lagrangian surface in
$\mathcal{\bar{M}}$. It is defined by the equations
\begin{equation}\label{Lagr-suface}
    \varphi^\ast_I=\frac{\partial \Psi}{\partial \varphi^I}\, , \qquad
    \text{gh} (\Psi)=1, \quad \varepsilon(\Psi)=1\,.
\end{equation}
The gauge fermion $\Psi$ involves the gauge fixing conditions
imposed on the original fields and ghosts of so called non-minimal
sector which are cohomologically trivial. The gauge invariance of
the path integral is guaranteed by its independence from the choice
of $\Psi$.

Let us first discuss the options of imposing the gauge fixing
conditions on the original fields, and then turn to introducing the
ghosts of non-minimal sector.

In principle, one can choose the gauge fixing conditions proceeding
from the interpretation of the BV formalism for the unfree gauge
theory as the BV theory of the equivalent gauge field theory of the
original fields and the compensators and with unconstrained gauge
parameters.  Then, the independent gauge fixing conditions
$\chi^\alpha(\phi,\xi)$ are imposed on the original fields and the
compensators
\begin{equation}\label{chi-xi-phi}
    \chi^\alpha(\eta)=0\, , \quad  A=(i,a), \quad \eta^A=(\phi^i;\xi^a)\, , \qquad
    \det\left(\Gamma^A_\alpha\frac{\partial \chi^\beta}{\partial
    \eta^A}\right){}_{\xi=0}\neq 0\, .
\end{equation}
The expression $\det\left(\Gamma^A_\alpha\frac{\partial
\chi^\beta}{\partial \eta^A}\right)$ is the FP determinant of the
gauge transformations (\ref{gtxi}), (\ref{Rxi}) evaluated at the
point where the compensator fields vanish. Notice that the gauge
transformations (\ref{gtxi}), (\ref{Rxi}) essentially affect the
compensator fields $\xi^a$ in the sense that the gauge orbits are
inevitably transverse to the level surface of the compensators
$\xi^a$. This is seen from the explicit form of the transformation
(\ref{gtxi}), (\ref{Rxi}) and irreducibility conditions for the
operators of the gauge parameter constraints (\ref{Constr-ired0}).
This means, the gauge conditions (\ref{chi-xi-phi}) implicitly fix
$\xi^a$, i.e. $\chi^\alpha(\phi,\xi)\approx 0\,\Rightarrow\,
\xi^a\approx f^a(\phi)$. This is much alike the gauge fixing in the
theory with the gauge invariance such that includes the
St\"uckelberg symmetry: the gauge conditions should fix the
St\"uckelberg fields, at least implicitly. To make the immediate
contact with the original theory without compensator fields, it
would be instructive to impose the gauge conditions explicitly
forcing $\xi^a$ to vanish. This can be done in two different ways,
at least.

The first option is to choose the gauge conditions
(\ref{chi-xi-phi}) explicitly separated into two subsets
$\chi^\alpha=(\chi^A(\phi),\xi^a) $:
\begin{equation}\label{chi-xi-0}
\alpha=(A,a)\,\quad \chi^A(\phi)=0\,, \quad\xi^a=0\,,
\end{equation}
where $\chi^A(\phi)$ are the independent gauge conditions imposed on
the original fields. The latter can be understood as the gauge
conditions of the original theory with unfree gauge symmetry
parameters. Notice that the number of the independent gauge
conditions $\chi^A(\phi)$ is less that the number of the generators
of unfree gauge symmetry transformations by the number of
constraints on the gauge parameters (\ref{eps-constr}). Because of
the numbers, the independent gauge conditions might be impossible to
choose in explicitly relativistic covariant way, even if the gauge
parameters are covariant. For example, if the gauge parameter is a
vector subjected to the scalar constraint, the number of independent
conditions should be $d-1$. There are no tensorial structures of
this dimension, while it seems unlikely to find $d-1$ independent
scalars such that could serve as the gauge conditions. That is why
it might be useful to consider redundant gauge fixing conditions if
the explicit covariance is the issue upon gauge fixing -- it is the
second option.

Given the independent gauge conditions (\ref{chi-xi-0}),
corresponding Lagrange multipliers, ghosts, and anti-fields read:
\begin{equation}\label{barCpi}
\text{gh}\pi_A =\text{gh}\pi_a
=0\,,\qquad\text{gh}\bar{C}_A=\text{gh}\bar{C}_a=-1\, ,\qquad
\text{gh}\bar{C}{}^{*A}=\text{gh}\bar{C}{}^{*a}=0 \, .
\end{equation}
The non-minimal action is constructed in the usual way,
\begin{equation}\label{S-non-min}
    S_{non-min}=S+ \bar{C}{}^{*A}\pi_A + \bar{C}{}^{*a}\pi_a \, .
\end{equation}
The gauge Fermion reads
\begin{equation}\label{Psi}
    \Psi= \bar{C}_A \chi^A(\phi) + \bar{C}_a \xi^a .
\end{equation}
This allows one to fix the anti-fields $\varphi^*$ reducing the
master action to the Lagrangian surface (\ref{Lagr-suface}):
\begin{equation}\label{SPsi}
    S_\Psi=S_{non-min}{}_{\big|\varphi*\mapsto \frac{\partial \Psi}{\partial
    \varphi}} \, .
\end{equation}
Given the gauge conditions $\xi^a=0$, the equations of the
Lagrangian surface (\ref{Lagr-suface}) include the relations
$\xi^\ast_a=\bar{C}_a$, so the anti-fields $\xi^\ast_a$ a replaced
in the gauge-fixed action $S_\Psi$ by the ghosts $\bar{C}_a$. Once
the gauge conditions (\ref{chi-xi-0}) kill off $\xi^a$, they can be
put to zero in the gauge fixed action. After these reductions, the
gauge fixed action reads
\begin{equation}\label{SchiA}
   S_{\chi} (\phi^i, C^\alpha, \bar{C}_A, \pi^A)=
   S_\Psi{}_{\big| \xi^a=0}=
   S(\phi) +\pi_A\chi^A(\phi)+\bar{C}_A\Gamma^i_\alpha \frac{\partial \chi^A}{\partial \phi^i} C^\alpha +  \bar{C}_a\Gamma^a_\alpha
    C^\alpha+  \ldots \, .
\end{equation}
It can be considered as the most general BRST invariant action of
the theory with unfree gauge symmetry when the gauge is fixed by
imposing the independent conditions $\chi^A$ on the original fields.
The ghosts $C^\alpha$ are assigned to all the gauge parameters, even
though they are unfree. The anti-ghosts $\bar{C}_A$ are assigned to
the gauge fixing conditions, while the anti-ghosts $\bar{C}_a$ are
assigned to all the constraints on the gauge parameters
(\ref{eps-constr}). The first item in $S_{\chi}$ is the original
action, the second one fixes the gauge. The third  item is the FP
term, while the  fourth one can be considered as a product of the
constraints on the ghosts (\ref{gh-constr}) to the Lagrange
multiplier $\bar{C}_a$. The higher order terms can involve, in
principle, the squares and higher orders of $\bar{C}_a$, so it is
not a true multiplier.

The number of the independent gauge conditions $\chi^A(\phi)$
(\ref{chi-xi-0}) is less than the number of gauge parameters, so
they most likely would break explicit relativistic symmetry. If the
explicit Poincar\'e or AdS symmetry is needed, then the number of
gauge conditions should coincide with the number of gauge
parameters. Once the parameters are unfree, the gauge conditions
should be redundant. In a slightly different wording, in the
theories with unfree gauge invariance, the Coulomb-type gauge
conditions can be independent, while the Lorentz-type ones cannot.
With this regard, it is useful to consider the redundant gauge
conditions imposed on the original fields, and fix the compensators,
\begin{equation}\label{chi-xi}
    \chi^\alpha(\phi)=0\, , \quad \xi^a=0\,,
\end{equation}
The fact that the gauge conditions $\chi^\alpha(\phi)$ are redundant
means
\begin{equation}\label{chi-red}
\exists
    \Delta^a_\alpha(\phi): \,\,\Delta^a_\alpha\chi^\alpha\approx 0 \, .
\end{equation}
The null-vectors $\Delta$ are supposed irreducible,
\begin{equation}\label{M-red}
\exists
    \Lambda^\alpha_b\,:\quad\det{\left( M^a_b\right)}\neq 0\,, \quad  M^a_b= \Lambda^\alpha_b\Delta_\alpha^a\, .
\end{equation}
The redundant gauge conditions are usually imposed in the theories
with reducible gauge symmetries \cite{BV2}, \cite{BV3}. With the
unfree the non-minimal sector anti-ghosts are introduced for the
redundant gauges by the same scheme, though it is asymmetric with
respect to the ghosts of the minimal sector, unlike the reducible
gauge symmetry. Given the reducible gauge conditions, the
non-minimal sector read
\begin{equation}\label{barC-red1}
    \text{gh} \bar{C}_a=\text{gh}\bar{C}_\alpha=\text{gh}\zeta^*_a = \text{gh}\pi{}^{*\alpha}=\text{gh}\pi{}^{*a} =-\text{gh}\lambda^a=-1 ;\,
    \text{gh}\lambda^*_a=-2;\,
\end{equation}
\begin{equation}\label{barC-red2}
    \text{gh}\zeta^a = \text{gh}\pi_\alpha =\text{gh}\bar{C}{}^{*\alpha}=\text{gh}\bar{C}{}^{*a}=0 \, .
\end{equation}
Here, $\bar{C}_\alpha$ are the anti-ghosts to the redundant gauge
conditions $\chi^\alpha(\phi)$, anti-ghosts $\bar{C}_a$ correspond
to the conditions fixing $\xi^a$, $\pi_\alpha$ are the Lagrange
multipliers to the redundant gauge conditions, $\lambda^a$ are the
ghosts for the gauge symmetry of the Lagrange multipliers
$\pi_\alpha$, and $\zeta^a$ are the Lagrange multipliers to the
gauge conditions imposed on $\pi_\alpha$. All the non-minimal sector
variables are introduced with their anti-fields. In the master
action these variables are included in the same way as in the
theories with reducible gauge symmetries,
\begin{equation}\label{S-nonmin-red}
    S_{non-min}=S(\varphi,\varphi^\ast) + \bar{C}{}^{*\alpha}\pi_\alpha +
    \bar{C}{}^{*a}\pi_a +\lambda^a\zeta^*_a
\end{equation}
The gauge Fermion reads
\begin{equation}\label{Psi-red}
    \Psi=\bar{C}_\alpha(\chi^\alpha+\Lambda^\alpha_b\zeta^b
    )+\bar{C}_a\xi^a \, .
    \end{equation}
On the Lagrange surface (\ref{Lagr-suface}) defined by $\Psi$ all
the antifields are excluded from the master action. In particular,
the anti-fields $\xi^\ast_a$ are replaced by $\bar{C}_a$. The
conjugate fields $\xi^a$ are killed off by the gauge conditions
(\ref{chi-xi}). As a result, the gauge fixed action depends only on
the original fields, ghosts to the unfree gauge symmetry, and
extended set of the anti-ghosts:
\begin{eqnarray}
\nonumber
  S_\chi &=& S_{\text{non-min}}{}_{\big| \varphi^\ast=\frac{\partial\Psi}{\partial\varphi};\,\xi^a=0}=S(\phi)+\pi_\alpha\chi^\alpha(\phi)+ \zeta^b\Lambda_b^\alpha\pi_\alpha + \\
\label{Schi-red}  \phantom{S} &+& \bar{C}_b\Gamma^b_\alpha
    C^\alpha+ \bar{C}_\beta\Gamma^i_\alpha \frac{\partial \chi^\beta}{\partial \phi^i} C^\alpha +
        \lambda^b\Lambda_b^\alpha\bar{C}_\alpha +  \ldots \, .
\end{eqnarray}
The gauge fixed action begins with the original one. Then, it
involves two gauge fixing terms. The first one involves the
redundant gauge conditions $\chi^\alpha$, while the second item
fixes the gauge for the multipliers $\pi_\alpha$.  The second line
in (\ref{Schi-red}) includes the modification of the FP action term
for the theory with unfree gauge symmetry and redundant gauge
conditions.. If the gauge conditions $\chi^\alpha$ were explicitly
split into the independent ones, and the identical zeros
$\chi^\alpha=(\chi^A;\chi^a\equiv 0)$, the extra ghosts $\lambda^a$,
the corresponding multipliers $\pi_a$, and the variables $\zeta^a$
would decouple from the other terms, and their contributions would
cancel each other in the path integral. The remaining action would
coincide with the expression (\ref{SchiA}) corresponding to the
independent gauge conditions.

\section{The unique existence of solution to the master equation}
In Section 3, the master equation has been formulated for the
theories with unfree gauge symmetry algebra. We have explicitly
found the solution in the second order approximation with respect to
the resolution degree (\ref{S1}), (\ref{S2}). Now, we consider the
existence problem in any order. At first, we shall see that the
problem can be brought to the usual HPT setup with respect to the
Koszul-Tate differential (\ref{KT-p}). Then, we consider the
cohomology of $\delta$. The differential $\delta$ can be understood
a resolution for the ideal $\bar{I}$  (\ref{barI}) of the on-shell
vanishing local function(al)s. The ideal $\bar{I}$ is generated not
only by the left hand sides of the Lagrangian equations, but also by
the completion functions (\ref{tau-a}), (\ref{Completion}), and by
the constraints imposed on the ghosts (\ref{gh-constr}). Once the
generating set is different for $\bar{I}$ comparing to the
corresponding ideal in the case of the Lagrangian theory with
unconstrained gauge symmetry parameters, the cohomology of the
differential has to be examined in this case. We shall demonstrate
that $\delta$ is acyclic in the strictly positive resolution
degrees. Given the acyliclicity of $\delta$, and relation
(\ref{delta-gamma}), we shall see that the master equation can be
iteratively solved by the usual HPT tools.  And finally, we shall
address the issue of physical observables.

The existence problem reads: given the first two orders
\begin{equation}\label{S-bound}
    S_0(\varphi,\varphi^\ast)=S(\phi), \qquad
    S_1(\varphi,\varphi^\ast)=  \tau_a(\phi)\xi^a+(\phi^*_i\Gamma^i_\alpha(\phi) + \xi^*_a \Gamma^a_\alpha(\phi)
    )C^\alpha\,,
\end{equation}
in the expansion of the master action with respect to the resolution
degree
\begin{equation}\label{S-exp-KT}
    S(\varphi,\varphi^\ast)=\sum_{k=0}S_k\,, \qquad \text{gh}\,(S_k)\,=\,\varepsilon\,(S_k)\,=\,0\, , \qquad
    \text{deg}\,(S_k)\,=\,k\, ,
\end{equation}
to iteratively solve the master equation (\ref{SS}) for all the
higher orders $S_k,\, k>1$.

The solution goes along the usual lines of the BV method. We
substitute the expansion (\ref{S-exp-KT}) into the master equation
$(S,S)=0$, and expand the left hand side with respect to the
resolution degree. In zero order, the equation is obeyed because of
the modified Noether identities (\ref{GI}),  as is noticed in
Section 2, see (\ref{SS0}). In the the order $k\geq 1$, it has the
structure
\begin{equation}\label{KTS}
    \delta S_{k+1}= B_k\, , \qquad \text{gh}B_k=1\, \quad \text{deg}
    B_k=k \, ,
\end{equation}
where $B_k$ is constructed of the derivatives of $S_l, \, l\leq k$.
The solution will exist for $S_{k+1}$ if $B_k$ is $\delta$-exact.
Because of the Jacobi identity for the anti-bracket, one can see
that $B_k$ is $\delta$-closed:
\begin{equation}\label{deltaBk}
    \delta B_k=0\, .
\end{equation}
 So, the existence problem reduces to the issue of
 $\delta$-cohomology in the positive resolution degrees.

Let us demonstrate that the Koszul-Tate differential is acyclic in
the class of functions with  $\text{deg}>0$. For simplicity, we
consider the functions of the ghost number one, and resolution
degree one. The higher degrees are treated in a similar way, while
the manipulations are more lengthy. The most general $B^1_1$ reads
\begin{equation}\label{B_1^1}
\text{gh}( B^1_1)=1\,,\quad   \deg(
B^1_1)=1\quad\Leftrightarrow\quad B^1_1=C^\alpha B_{a\alpha}\xi^a+
\frac{1}{2}C^\alpha
    C^\beta\left(B^i_{\alpha\beta}\phi^*_i+
    B^i_{\alpha\beta}\xi^*_a\right).
\end{equation}
Suppose $B^1_1$ is $\delta$-closed. Then, the structure functions
$B_{a\alpha}(\phi),, B^i_{\alpha\beta}(\phi)$ obey the relations :
\begin{equation}\label{deltaB^1_1}
\delta B^1_1=C^\alpha C^\beta\left( \Gamma^a_\alpha B_{a\beta}
+B^i_{\alpha\beta}\partial_i S + B^a_{\alpha\beta}\tau_a \right)=0\,
.
\end{equation}
These relations involve the left hand sides of the Lagrangian
equations, the completion functions (\ref{tau-a}), and the operators
of gauge parameter constraints (\ref{eps-constr}). These quantities
are involved in the modified Noether identities (\ref{GI}), and they
are assumed to obey the regularity and completeness conditions, see
(\ref{Completion}), (\ref{G-complete}), (\ref{GT-irred}). These
relations have the consequences (\ref{Constr-ired0}),
(\ref{Constr-irred1}), (\ref{Constr-irred2}), (\ref{Constr-irred3}).
Given the regularity and completeness assumptions for the unfree
gauge symmetry algebra, and their consequences, the relation
(\ref{deltaB^1_1}) means
\begin{equation}\label{existsAU}
\exists A_{(ab)}, \, A^i_{b\alpha},\, A^a_{b\alpha},\,
U_{\alpha\beta}^\gamma,\, E^{[ij]}_{\alpha\beta},\,
E^{ia}_{\alpha\beta},\, E^{[ab]}_{\alpha\beta}:
\end{equation}
\begin{equation}\label{B1}
    B_{a\alpha}= A_{ab}\Gamma^b_\alpha + A^i_{a\alpha}\partial_i S +
    A^b_{ a \alpha}\tau_b\, ;
\end{equation}
\begin{equation}\label{B2}
    B^i_{\alpha\beta}=
\Gamma^a_{[\alpha } A^i_{\beta]a} + U_{\alpha\beta}^\gamma
\Gamma_\gamma^i+ E^{[ij]}_{\alpha\beta}\partial_j S +
E^{[ia]}_{\alpha\beta}\tau_a\, ;
\end{equation}
\begin{equation}\label{B3}
    B^a_{\alpha\beta}=
\Gamma^b_{[\alpha } A^a_{\beta]b} + U_{\alpha\beta}^\gamma
\Gamma_\gamma^a - E^{aj}_{\alpha\beta}\partial_j S +
E^{[ab]}_{\alpha\beta}\tau_b\,.
\end{equation}
Once the structure coefficients
$B_{a\alpha},\,B^i_{\alpha\beta},\,B^a_{\alpha\beta}$ read as above,
$B^1_1$ is $\delta$-exact:
\begin{equation}\label{BdeltaD}
B^1_1=\delta D^0_2\,,\qquad\text{gh}D^0_2=0,\quad\varepsilon
D^0_2=0,\quad \text{deg}D^0_2=2\,,
\end{equation}
where
\begin{equation}\label{D^0_2}
 D^0_2=\frac{1}{2}A_{ab}\xi^a\xi^b + C^\alpha \xi^b \left(\phi^*_i A^i_{b\alpha}+ A^a_{b\alpha}\xi^*_a\right)
 + \frac12 C^\alpha C^\beta\left( U_{\alpha\beta}^\gamma C^*_\gamma+
E^{ij}_{\alpha\beta}\phi^*_i\phi^*_j + E^{ia}_{\alpha\beta}
\phi^*_i\xi^*_a + E^{ab}_{\alpha\beta}\xi^*_a\xi^*_b\right) \, .
\end{equation}
This means, there exists $S_2=D^0_2+\delta\Psi, \, \text{gh}\Psi=-1,
\, \text{deg}\Psi=3$. Once $\delta$ is acyclic in the higher
resolution degrees, this proves the unique\footnote{Up to the
$\delta$-exact terms. This is a natural ambiguity related to the
fact, that the structure functions of the gauge algebra are defined
modulo the on shell vanishing contributions.} existence of all the
higher order items in the expansion (\ref{S-exp-KT}).

Let now discuss the details of construction of the BRST invariant
physical observables by means of the HPT method. The problem reads
as follows: given the  first two orders
\begin{equation}\label{O-O1-brst}
    \mathcal{O}_0=O(\phi)\,,\qquad
    \mathcal{O}_1=C^\alpha(V_\alpha^i\phi^\ast_i+V^a_\alpha\xi^\ast_a)+W_a\xi^a\,,
\end{equation}
in the expansion of the physical observable in the resolution degree
(\ref{O-brst}), to iteratively solve equation (\ref{sO-exp}) for all
the higher orders $\mathcal{O}_k, k>2$.

The solution goes along the usual lines of the BV method, given the
acyclicity of $\delta$ in positive resolution degrees. For $k=1$
equation (\ref{sO-exp}) is satisfied because of the condition
(\ref{gammaO}). For $k>1$, the structure of equation (65) reads
\begin{equation}\label{}
    \delta \mathcal{O}_{k+1}=\mathcal{B}_k,\qquad \text{gh}\,
    \mathcal{B}_k=1\,,\qquad \text{deg}\,\mathcal{B}_k=k\,,
\end{equation}
where $B_k$ is constructed of the derivatives of $\mathcal{O}_l, \,
l\leq k$. The solution will exist for $\mathcal{O}_{k+1}$ if
$\mathcal{B}_k$ is $\delta$-exact. Because of the Jacobi identity
for the anti-bracket, one can see that $\mathcal{B}_k$ is
$\delta$-closed:
\begin{equation}\label{deltaBk}
    \delta \mathcal{B}_k=0\, .
\end{equation}
The quantity $\mathcal{B}_k$ is $\delta$-exact because the
Koszul-Tate differential $\delta$ is acyclic in any positive
resolution degree. This proves the existence of the BRST invariant
physical observables. As in the usual BV formalism, the BRST
invariant is unique for any original physical observable, modulo
BRST-exact terms. So the factor algebra $R/I$ of the nontrivial
gauge invariant function(al)s is isomorphic to the BRST cohomology
group of zero ghost number.

\section{Examples}

The best known example of unfree gauge symmetry is provided by the
volume preserving diffeomorphism, $T$-diff.  $T$-diff is the gauge
symmetry of unimodular gravity and various modifications, see
\cite{Buchmuller}, \cite{Unruh:1988in}, \cite{Henneaux},
\cite{Ellis1}, \cite{Ellis2}, \cite{Padilla:2014yea},
\cite{Gielen:2018pvk}, \cite{Kamenshchik}, \cite{Barvinsky} and
references therein.  Once the unimodularity condition  $\det
g_{\mu\nu}=-1$ is imposed, the gauge variation $\delta_\epsilon
g_{\mu\nu}=\nabla_\mu\epsilon_\nu+\nabla_\nu\epsilon_\mu \,$ is
unfree of the metric, as the transformation have to be consistent
with the fixed volume. This means, the transformation parameter
$\epsilon^\mu(x)$ is constrained by the transversality condition,
\begin{equation}\label{xi-t}
\delta_\epsilon \det{g_{\mu\nu}}=0\quad\Leftrightarrow\quad
\partial_\mu\epsilon^\mu=0 \, .
\end{equation}
A similar phenomenon is known in the model of the massless free spin
2 described by traceless tensor field \cite{Alvarez:2006uu},
\cite{Blas:2007pp}. In this case, the Fierz-Pauli gauge symmetry
parameter has to obey the transversality condition (\ref{xi-t}). The
extension is known of this class of the massless spin models to the
fields of any spin $s$ \cite{Skvortsov:2007kz}. The gauge
parameters, being the symmetric traceless tensors of rank $s-1$ turn
out unfree again. The Maxwell-like models \cite{Campoleoni:2012th},
\cite{Francia:2016weg} of higher spins involve tracefull tensors,
whose gauge symmetry is parameterized by the tracefull tensors of a
lower rank, while the differential equations  are still imposed on
the gauge parameters. So,  the phenomenon of the unfree gauge
symmetry is not necessarily related to any constraint (like trace
condition) imposed on the fields.

In all the above cases, the regularity assumption (\ref{T}) is
invalid, while the relaxed conditions (\ref{tau-a}),
(\ref{Completion}) hold true, at least modulo some subtleties
related to global degrees of freedom. In the model of spin 2
\cite{Alvarez:2006uu}, \cite{Blas:2007pp}, the field equations for
the Minkowski space traceless tensor $h^{\mu\nu}$ has a differential
consequence:
\begin{equation}\label{dddh}
    \partial_\lambda\partial_\mu\partial_\nu h^{\mu\nu}\approx 0\, .
\end{equation}
This means, $\partial_\mu\partial_\nu h^{\mu\nu}\approx const$. The
usual boundary conditions imply the fields to vanish on the spacial
infinity in Minkowski space.\footnote{The boundary conditions should
not be confused with Cauchy data which define the initial values of
the fields and their time derivatives on some space-like
hyper-surface, not at spacial infinity. Zero boundary conditions do
not contradict to nontrivial Cauchy data.} With zero boundary
conditions, the constant should vanish, so we have
\begin{equation}\label{ddh}
\tau\equiv\partial_\mu\partial_\nu h^{\mu\nu}\approx 0\, .
\end{equation}
The quantity $\tau$ is a function of derivatives of the fields, it
is local. It vanishes on-shell, given the boundary conditions.
However, the relation $\tau\approx 0$ is not a differential
consequence of the Lagrangian equations. So, we have the on-shell
trivial local quantity which does not reduce to a linear combination
of the Lagrangian equations and their derivatives. It should be
considered as a completion function as defined by relations
(\ref{tau-a}). Notice that the higher spin extensions
\cite{Skvortsov:2007kz} of the spin two traceless tensor field
theory \cite{Alvarez:2006uu}, \cite{Blas:2007pp} also reveal a
existence of the completion functions corresponding to the
definition (\ref{tau-a}). The Maxwell-like models of higher spins
\cite{Campoleoni:2012th}, \cite{Francia:2016weg}  also have the
similar local quantities such that vanish on shell and do not reduce
to the linear combinations of the Lagrangian equations. Detailed
explanations of these quantities can be found in the reference
\cite{Francia:2016weg}. In the unimodular gravity, the field
equations have the differential consequence
\begin{equation}\label{dR}
    \partial_\mu R\approx 0\,\quad \Rightarrow \quad
    R\approx\Lambda\, ,\quad \Lambda=const\,,
\end{equation}
where $R$ is the Ricci curvature of the unimodular metric. This
means,
\begin{equation}\label{R}
    R\approx\Lambda\, ,\quad \Lambda=const\,,
\end{equation}
If the space is supposed to be asymptotically flat, $\Lambda$ has to
be zero. If the metric has the (A)dS  asymptotics, $\Lambda$ is a
fixed constant being defined by the asymptotic (A)dS curvature
radius. So, with any fixed asymptotics of the metric, the completion
function (\ref{tau-a}) would be $\tau=R-\Lambda$. If the assymtotics
is not fixed of the metrics, then the constant $\Lambda$ can be
considered as a modular parameter involved in the same completion
function $\tau=R-\Lambda$. The subtleties concerning the account for
the modular parameters are beyond the scope of this paper. We can
only notice, that once the completion function is explicitly
involved in the formalism, the presence of the modular parameters is
also made explicit that could facilitate the study of their impact
on the dynamics.

\vspace{0.2 cm}

Now, let us consider the specific model to exemplify the general
formalism. The linearized unimodular gravity is theory of
second-rank symmetric traceless tensor field $h_{\mu\nu}(x)$ with
the action functional
\begin{equation}\label{L}
    S_0=\int L d^4x\,,\qquad
    L=\frac{1}{2}\bigg(\partial_\mu h_{\mu\nu}\partial^\mu
    h^{\nu\rho}-2\partial^\mu h_{\mu\rho}\partial_\nu
    h^{\nu\rho}\bigg)\,,\qquad h^\mu{}_\mu\equiv0\,.
\end{equation}
The gauge symmetry of the action is the linearized volume-preserving
diffeomorphism T-diff,
\begin{equation}\label{T-diff}
    \phantom{\frac12}\delta_\epsilon h^{\mu\nu}=\partial^{\mu}\epsilon^\nu+\partial^\nu\epsilon^{\mu}\,,
    \qquad \partial_\mu\epsilon^\mu=0\,,\phantom{\sum}
\end{equation}
where $\epsilon^\mu$ is the transformation parameter. The gauge
symmetry is constrained because the transformation parameter has to
obey the transversality condition.

The Lagrangian equations  read:
\begin{equation}\label{LE-uni}
    \frac{\delta S}{\delta h^{\mu\nu}}=-\bigg(\Box
    h_{\mu\nu}-\partial_\mu\partial^\rho h_{\rho\nu}-\partial_\nu\partial^\rho
    h_{\rho\mu}+\frac{1}{2}\partial_{\rho}\partial_{\sigma}h^{\rho\sigma}\bigg)\approx0\,,\,.
\end{equation}
Taking the divergence of the equations, we arrive at the consequence
$\partial_\mu \tau\approx 0$, where $\tau=\partial_\mu\partial_\nu
h^{\mu\nu}$. Given  the boundary conditions that the fields are
vanishing at the space infinity, we arrive at (\ref{ddh}). The
completion function $\tau$ is not a linear combination of the right
hand sides of the Lagrangian equations.  One can see that the
completion function is preserved by the $T$-diffs,
\begin{equation}\label{delta-tau-uni}
    \phantom{\frac12}\delta_\epsilon\tau=2\Box\partial_\mu\epsilon^\mu\,.\phantom{\frac12}
\end{equation}
The gauge invariance of the completion function holds true only with
account of the constraint on gauge parameter, so the structure
function $W$ (\ref{Gtau}), (\ref{W-symm}) does not vanish. In the
case at hands
\begin{equation}\label{W-uni}
    W=2\Box\,.
\end{equation}
This structure function is symmetric (cf. (\ref{W-symm})), because
the D'Alembert operator is formally self-adjoint, as it should be.

There is the modified Noether identity (\ref{GI}) involving the
Lagrangian equations (\ref{LE-uni}) and completion function
(\ref{ddh}),
\begin{equation}\label{GI-uni}
    \partial^\nu \frac{\delta S}{\delta
    h^{\mu\nu}}-\frac{1}{2}\partial^\mu \tau\equiv0\,.
\end{equation}
The gauge identity generators have the form
\begin{equation}\label{GG-uni}
    \phantom{\sum}\Gamma_\rho^{\mu\nu}=\delta^\mu{}_\rho\partial^\nu+\delta^\mu{}_\rho\partial^\nu\,,\qquad
    \Gamma_\rho=-\partial_\rho\,.\phantom{\sum}
\end{equation}
As we see, these quantities determine the gauge transformation
(\ref{T-diff}) of the theory (\ref{L}) and constraint on parameter
by the rule (\ref{GT}), (\ref{eps-constr}). The gauge generators are
field independent, so the structure functions $R^i_{\alpha a},
R^{b}_{\alpha a}, E^{ij}_{\alpha\beta}, E^{ia}_{\alpha\beta},
E^{ab}_{\alpha\beta}$ (\ref{GiGi}), (\ref{GiGa}) vanish identically.

To construct the BV-BRST embedding of the linearized unimodular
gravity, we introduce the ghosts and anti-fields following the
general prescription (cf. Table 1) :
\begin{equation}\label{f-af-uni}
    \phantom{\frac12}h^{\mu\nu}(x)\,,
    \qquad h^\ast{}_{\mu\nu}(x)\,,
    \qquad C^\mu(x)\,,
    \qquad C^\ast{}_\mu(x)\,,
    \qquad \xi(x)\,,
    \qquad \xi^\ast(x)\,.\phantom{\frac12}
\end{equation}
The condensed indices labeling the fields $\phi^i$, gauge parameters
$\epsilon^\alpha$, and the completion functions $\tau_a$ are
uncondensed in the following way: $i=\{\mu\nu,x\}$;
$\alpha=\{\mu,x\}$, $a=\{x\}$, where $x$ is the space-time argument.
The gradings of the fields and anti-fields are arranged in Table 2.
\begin{center}
\begin{tabular}{|c|c|c|c|c|c|c|}
  \hline
  grading$\backslash$variable & $\phantom{0}h^{\mu\nu}$ \phantom{0}& \phantom{0} $\xi$ \phantom{0}& \phantom{0} $C^\mu$ \phantom{0} &
  \phantom{0} $h^*_{\mu\nu}$ \phantom{0} & \phantom{0}$\xi^*$ \phantom{0}& \phantom{0} $C^*_\mu$ \phantom{0} \\
  \hline
  $\varepsilon $ & 0 & 0 & 1 &  1  &  1&  0\\
  $\text{gh}$        & 0 & 0 & 1 & -1  & -1& -2\\
%$\text{pgh}$       & 0 & $\textbf{1}$ & 1 &  0  &  0&  0 \\
  $\text{deg}$       & 0 & 1 & 0 &  1  &  1&  2 \\
  \hline
\end{tabular}
\end{center}
\begin{center}
Table 2.
\end{center}
The mater action  (\ref{SBV-exp}) has the following form:
\begin{equation}\label{S-min-uni}
    \phantom{\frac12}S=\int \bigg(L+\xi\,\partial_\mu\partial_\nu h^{\mu\nu}-\partial_\mu\xi\partial^\mu\xi+
    h^\ast{}_{\mu\nu}(\partial^\mu C^\nu+\partial^\nu C^\mu)-\xi^\ast\partial_\mu C^\mu
    \bigg) d^4x\,.\phantom{\frac12}
\end{equation}
In this expression, the contributions of the resolution degree 0, 1,
2 are determined by formulas (\ref{S0}), (\ref{S1}), (\ref{S2}),
with the competition function $\tau$ and structure functions
$\Gamma^{\mu\nu}{}_\rho,\,\Gamma_\rho,\,W$ being defined by the
relations (\ref{ddh}), (\ref{GG-uni}), (\ref{W-uni}). All the
contributions of resolution degrees vanish identically,
$S_k=0,k\geq3$.

The gauge fixing conditions  can be introduced in various ways. The
first option is to impose the independent gauge fixing conditions
(\ref{chi-xi-0}), which we chose in the form
\begin{equation}\label{chi-xi-0-uni}
    \chi^i(h)=\partial_j h^{ij}-\frac{1}{2}\partial^i
    h^j{}_j=0\,,\qquad \xi=0\,.
\end{equation}
Here, the quantity $h^{ij}, i,j=1,2,3,$ denotes the space part of
the symmetric tensor $h^{\mu\nu}$. The gauge is not covariant
because it is not possible to find the tensor structure with three
independent components in four-dimensional Minkowski space.

Given the independent gauge conditions (\ref{chi-xi-0-uni}), the
corresponding Lagrange multipliers, ghosts, and anti-fields read:
\begin{equation}\label{barCpi-uni}
\text{gh}\pi_i(x) =\text{gh}\pi (x) =0\,,\qquad\text{gh}\bar{C}_i
(x)=\text{gh}\bar{C} (x)=-1\, ,\qquad
\text{gh}\bar{C}{}^{*i}(x)=\text{gh}\bar{C}{}^{*}(x)=0 \, .
\end{equation}
The non-minimal action (\ref{S-non-min}) takes the following form:
\begin{equation}\label{S-non-min-uni}
    S_{non-min}=S+\int \Big(\bar{C}{}^{*i}\pi_i + \bar{C}{}^{*}\pi \Big)d^4 x\, .
\end{equation}
The gauge Fermion (\ref{Psi}) reads
\begin{equation}\label{Psi-uni}
    \Psi= \int \Big[\bar{C}_i \Big(\partial_j h^{ij}-\frac{1}{2}\partial^i
    h^j{}_j\Big) + \bar{C}\, \xi\Big]d^4 x .
\end{equation}
The gauge fixing conditions for anti-fields have the form
\begin{equation}\label{gf-anti-fields-uni}\begin{array}{c}\displaystyle
    h^\ast{}_{ij}=-\frac{1}{2}(\partial_i \bar{C}_j+\partial_j
    \bar{C}_i-\delta_{ij}\partial_k\bar{C}^k)\,,\qquad
    h^\ast{}_{0i}=0\,,\qquad \xi^\ast=\bar{C}\,,\\[5mm]\displaystyle
    \bar{C}{}^{*i}=\partial_j h^{ij}-\frac{1}{2}\partial^i
    h^j{}_j\,,\qquad \bar{C}{}^{*}=\xi\,.
\end{array}\end{equation}
The gauge fixed action (\ref{SchiA}) reads
\begin{equation}\label{S-psi-uni}
    S_\Psi=\int \bigg[L+\xi\,\partial_\mu\partial_\nu h^{\mu\nu}-\partial_\mu\xi\partial^\mu\xi+
    \bar{C}_i\Delta C^i-\bar{C}\partial_\mu C^\mu+\pi_i\Big(\partial_j h^{ij}-\frac{1}{2}\partial^i
    h^j{}_j\Big)+\pi\xi\bigg] d^4x\,.\phantom{\frac12}
\end{equation}
The generating functional of Green's functions is determined in the
usual way,
\begin{equation}\label{Z-general}
    Z=\int
    [d\varphi]\exp\bigg(\frac{i}{\hbar}S_\Psi\bigg)\,,\qquad
    \varphi=\{h^{\mu\nu}(x),C^{\mu}(x),\xi(x),\bar{C}_{i}(x),\bar{C}(x),\pi_i(x),\pi(x)\}\,.
\end{equation}
If the compensator field $\xi$ and Lagrange multiplier $\pi$ are
integrated out of the path integral, we get the following result:
\begin{equation}\label{Z-uni}\begin{array}{c}\displaystyle
    Z=\int
    [d\varphi']\exp\bigg\{\frac{i}{\hbar}\int \bigg[L+\bar{C}_i\Delta C^i-\bar{C}\partial_\mu C^\mu+\pi_i\Big(\partial_j h^{ij}-\frac{1}{2}\partial^i
    h^j{}_j\Big)\bigg] d^4x\bigg\}\,,\\[7mm]\displaystyle
    \varphi'=\{h^{\mu\nu}(x),C^{\mu}(x),\bar{C}_{i}(x),\bar{C}(x),\pi_i(x)\}\,.
\end{array}\end{equation}
This expression has been previously obtained in the work
\cite{Kaparulin:2019quz} by means of the modified FP method for the
theories with unfree gauge symmetry.

Consider another gauge fixing. Introduce covariant gauge fixing
condition of the following form:
\begin{equation}\label{gf-red-uni}
    \phantom{\frac12}\chi^\mu(h)=\partial_\nu h^{\mu\mu}=0\,,\qquad \xi=0\,.\phantom{\frac12}
\end{equation}
This gauge is redundant because the gauge conditions are not
independent on the mass shell,
\begin{equation}\label{gf-id-uni}
    \phantom{\frac12}\partial_\mu \chi^\mu=\partial_\mu\partial_\nu h^{\mu\mu}\approx0\,.\phantom{\frac12}
\end{equation}
The operator of identity (\ref{chi-red}) is just a divergence,
\begin{equation}\label{Delta-uni}
    \phantom{\frac12}\Delta^a_\alpha\equiv\Delta_\mu=\partial_\mu\,.\phantom{\frac12}
\end{equation}
The null vectors of $\Delta_\mu$ are irreducible in the sence
(\ref{M-red}), with $\Lambda$ being chosen as follows:
\begin{equation}\label{Lambda-uni}
    \phantom{\frac12}\Lambda_a^\alpha\equiv\Lambda^\mu=\partial^\mu\,.\phantom{\frac12}
\end{equation}
In this case, $M$ (\ref{M-red}) is the D'Alembert operator,
\begin{equation}\label{M-red-uni}
    \phantom{\frac12}\Lambda^\mu\Delta_\mu=\Box\,,\qquad \det\Box\neq0\,.\phantom{\frac12}
\end{equation}
The set of fields and anti-fields (\ref{barC-red1}),
(\ref{barC-red2}) of non-minimal sector read
\begin{equation}\label{anti-red-uni}\begin{array}{c}\displaystyle
    \phantom{\frac12} \text{gh}\, \bar{C}_\mu(x)\,=\,\text{gh}\,\bar{C}(x)\,
    =\,\text{gh}\,\zeta^*(x)\,=\,\text{gh}\,\pi{}^{*\mu}(x)\,=\,\text{gh}\,\pi{}^{*}(x)
    \,=\,-\text{gh}\,\lambda(x)=-1;\quad\,\text{gh}\,\lambda^*(x)=-2;\,\\[5mm]\displaystyle
    \phantom{\frac12}\text{gh}\,\zeta(x)\,=\,\text{gh}\,\pi_\mu(x)\,=\,\text{gh}\,\bar{C}{}^{*\mu}(x)
    \,=\,\text{gh}\,\bar{C}{}^{*}(x)=0 \, .
\end{array}\end{equation}
The associated non-minimal master-action (\ref{S-nonmin-red}) has
the form,
\begin{equation}\label{S-nonmin-red-uni}
    \phantom{\frac12}S_{nom-min}\,=\,S\,+\,\int
    \bigg(\bar{C}{}^{\ast\mu}\pi_\mu+\bar{C}{}^\ast\pi+\zeta^\ast\lambda\bigg)d^4x\,.\phantom{\frac12}
\end{equation}
The gauge Fermion (\ref{Psi-red}) reads
\begin{equation}\label{Psi-red-uni}
    \phantom{\frac12}
    \Psi=\int \bigg(\bar{C}{}_\mu\,(\partial_\nu h^{\mu\nu}\,+\,\partial{}^\mu\zeta)\,+\,\bar{C}\xi\bigg)d^4x\,.
\end{equation}
The anti-fields express as follows:
\begin{equation}\label{gf-anti-fields-red-uni}\begin{array}{c}\displaystyle
    h^\ast{}_{\mu\nu}=-\frac{1}{2}(\partial_{\mu}\bar{C}_\nu+\partial_\nu \bar{C}_\mu)\,,\qquad \xi^\ast=\bar{C}\,,\qquad
    \bar{C}{}^{*\mu}=\partial_\nu h^{\nu\mu}+\partial^\mu\zeta\,,\\[5mm]\displaystyle
    \bar{C}{}^{*}=\xi\,,\qquad
    \zeta^\ast=-\partial^{\nu}\bar{C}_\nu\,.
\end{array}\end{equation}
For the gauge fixed action (\ref{Schi-red}), we get
\begin{equation}\label{S-psi-red-uni}\begin{array}{c}\displaystyle
    S_\Psi=\int \bigg[L+\xi\,\partial_\mu\partial_\nu
    h^{\mu\nu}-\partial_\mu\xi\partial^\mu\xi-
    \partial_\mu \bar{C}_\nu(\partial^\mu C^\mu+\partial^\mu C^\nu)-\bar{C}\partial_\mu
    C^\mu+\\[5mm]\displaystyle
    \pi_\mu(\partial_\nu h^{\mu\nu}+\partial^\mu\zeta)+\pi\xi-
    \partial^{\mu}\bar{C}_\mu\lambda\bigg] d^4x\,.
\end{array}\end{equation} The generating functional of Green's functions
read
\begin{equation}\label{Z-red-general}\begin{array}{c}\displaystyle
    Z=\int
    [d\varphi]\exp\bigg(\frac{i}{\hbar}S_\Psi\bigg)\,,\quad
    \varphi=\{h^{\mu\nu}(x),C^{\mu}(x),\xi(x),\bar{C}_{\mu}(x),\bar{C}(x),
    \pi_\mu(x),\pi(x),\zeta(x),\lambda(x)\}\,.
\end{array}\end{equation}

Let us see that the expressions (\ref{Z-general}) and
(\ref{Z-red-general}) define one and same quantity $Z$. To make
explicit comparison of results with independent and redundant gauge,
we bring (\ref{Z-red-general}) to the form (\ref{Z-uni}). We proceed
with making the change of ghost variables,
\begin{equation}\label{ghost-uni-change}\begin{array}{c}\displaystyle
    \bar{C}_0=\bar{C}'{}_0-\partial_0\partial^iC'{}_i\,,\qquad
    \bar{C}_i=\partial_0\partial^0
    \bar{C}'{}_i\,,\qquad \pi_0=\pi'{}_0-\partial_0\partial^i\pi'{}_i\,,\qquad
    \pi_i=\partial_0\partial^0
    \pi'{}_i\,,\\[5mm]\displaystyle
    \partial_0\zeta=\zeta'\,,\qquad \partial_0\lambda=\lambda'\,.
\end{array}\end{equation}
This change of the variables preserves the integration measure
because the variables of opposite Grassmann parity transform with
one and the same transformation law. The intermediate result for the
path integral reads
\begin{equation}\label{Z-red-general-2}\begin{array}{c}\displaystyle
    Z=\int
    [d\varphi']\exp\bigg(\frac{i}{\hbar}\int \bigg[L+\xi\,\partial_\mu\partial_\nu h^{\mu\nu}-\partial_\mu\xi \partial^\mu\xi+\bar{C}'{}_i(\partial_0\partial^0\Box C^i
    +\partial_0\partial^i \Box C^0)-\bar{C}\partial_\mu
    C^\mu+\\[5mm]\displaystyle
    \bar{C}'{}_0(\partial_\mu\partial^\mu C^0+\partial^0\partial_\mu C^\mu)+\pi'{}_i(\partial_0\partial^0\partial_\mu h^{\mu i}
    -\partial^i\partial_0\partial_\mu h^{\mu 0})+\pi'{}_0(\zeta'+\partial_\mu h^{\mu0})+\bar{C}'{}_0\lambda'+\pi\xi\bigg]
    d^4x\bigg)\,,\\[6mm]\displaystyle
    \varphi'=\{h^{\mu\nu}(x), \xi(x), C^{\mu}(x),\bar{C}'{}_{\mu}(x),\bar{C}(x),
    \pi'{}_\mu(x),\pi(x),\zeta'(x),\lambda'(x)\}\,.
\end{array}\end{equation}
In this expression, the variables
$\zeta'\,,\lambda'\,,\pi'{}_0\,,\bar{C}'{}_0$ can be integrated out,
\begin{equation}\label{}\begin{array}{c}\displaystyle
    \int [d\pi'{}_0][d\bar{C}'{}_0][d\,\zeta'][d\,\lambda']
    \exp\bigg(\frac{i}{\hbar}\int \bigg[\pi'{}_0(\zeta'+\partial_\mu
    h^{\mu0})+\bar{C}'{}_0(\lambda'+\partial_\mu\partial^\mu C^0+\partial^0\partial_\mu C^\mu)\bigg]d^4x\bigg)=\\[5mm]\displaystyle=\int
    [d\pi'{}_0][d\bar{C}'{}_0]\delta(\pi'{}_0)\delta(\bar{C}'{}_0)
    \exp\bigg(\frac{i}{\hbar}\int \bigg[\pi'{}_0\partial_\mu
    h^{\mu0}+\bar{C}'{}_0(\partial_\mu\partial^\mu C^0+\partial^0\partial_\mu C^\mu)\bigg]d^4x\bigg)=\text{const}\,.
\end{array}\end{equation}
After simplifications, we get
\begin{equation}\label{Z-red-general-final}\begin{array}{c}\displaystyle
    Z=\int
    [d\varphi'']\exp\bigg(\frac{i}{\hbar}\int \bigg[L+\xi\,\partial_\mu\partial_\nu h^{\mu\nu}-\partial_\mu\xi \partial^\mu\xi+\bar{C}'{}_i(\partial_0\partial^0\Box C^i
    +\partial_0\partial^i \Box C^0)-\bar{C}\partial_\mu
    C^\mu+\\[5mm]\displaystyle+\pi'{}_i(\partial_0\partial^0\partial_\mu h^{\mu i}-\partial^i\partial_0\partial_\mu h^{\mu 0})+\pi\xi\bigg]
    d^4x\bigg)\,,\quad
    \varphi''=\{h^{\mu\nu}(x),\xi, C^{\mu}(x),\bar{C}'{}_{i}(x),\bar{C}(x),\pi'{}_i(x),\pi(x)\}\,.
\end{array}\end{equation}
This expression can be obtained from the gauge fixed action
(\ref{SchiA}) in the independent gauge,
\begin{equation}\label{Z-red-uni-final}
    \chi'{}^i(h)=\partial_0\partial^0\partial_\mu h^{\mu i}-\partial^i\partial_0\partial_\mu h^{\mu
    0}\,,\qquad \xi=0\,.
\end{equation}
As the path integral is independent of the gauge choice, the
expressions (\ref{Z-uni}) and (\ref{Z-red-uni-final}) determine one
and the same expression for the generating functional of Green's
functions of the linearized unimodular gravity.

\section{Concluding remarks}
Let us summarize and discuss the results.

Proceeding from the observation that the reasonable gauge field
theories can admit the local quantities termed the completion
functions such that vanish on shell and do not reduce to the
differential consequences of equations of motion (\ref{tau-a}), we
deduce the most general gauge symmetry algebra for this case. It
turns out that the existence of the completion functions in the
theory results in the unfree gauge symmetry, with gauge parameters
constrained by the equations (\ref{eps-constr}). And vice versa, the
unfree gauge symmetry implies the existence of completion functions.
This is a consequence of the modified Noether identities (\ref{GI})
which involve both the Lagrangian equations and completion
functions. The modified identities result in the unfree gauge
algebra. Given the unfree gauge symmetry algebra, we work out a
systematic procedure for the BV-BRST embedding of the theory. The
extension of the BV formalism to the systems with unfree gauge
symmetry algebra has some distinctions from the case where the gauge
parameters are unconstrained. The source of distinctions is
two-fold. First, the Koszul-Tate resolution for the ideal of
on-shell vanishing local function(al)s $I$ should involve the
completion functions as the l.h.s. of the Lagrangian equations do
not generate $I$. Second, the ghost are constrained by the equations
(\ref{gh-constr}) as the corresponding gauge parameters are unfree
(\ref{eps-constr}). The equations constraining the ghosts
(\ref{gh-constr}) have to be also accounted for by the Koszul-Tate
resolution. These reasons define the minimal set of the fields and
anti-fields needed for the proper BV embedding, see Table 1 in
Section 3. The set involves the anti-fields $\xi^a$ to the
constraints imposed on the ghosts (\ref{gh-constr}). What is
unusual, these anti-fields have the ghost number zero, while their
resolution degree is 1. The completion functions are also assigned
with the anti-fields $\xi^*_a$. The usual BV formalism with
unconstrained gauge symmetry parameters does not involve
$\xi,\xi^*$. These new variables are naturally conjugate with
respect to the anti-bracket. The boundary conditions for the BV
master-action are defined by the original action, gauge generators,
completion functions, and operators of gauge parameter constraints
(\ref{S0}), (\ref{S1}). In the case of the gauge symmetry with
unconstrained gauge parameters, only first two constituents are
involved. Given the regularity conditions imposed on the boundary,
the master equations admits a solution which is unique modulo the
natural ambiguity. The BV formalism for the unfree gauge symmetry
admits the re-interpretation in terms of the usual BV formalism for
the theory with the ``compensator fields'', see the remark in the
end of Section 3. Given the master-action, $\xi$ can be considered
on an equal footing with the original gauge fields $\phi$. Then, the
theory would correspond to the theory of the fields $\phi,\xi$ with
the action $S'(\phi,\xi)$ (\ref{S-xiexp}) and unconstrained gauge
symmetry of the extended set of fields. From the viewpoint of this
re-interpretation, the existence theorem of Section 5 provides a
systematic procedure for inclusion of the compensator fields such
that the gauge symmetry involves unconstrained parameters of the
extended theory. The equivalence  is obvious between the original
theory and theory with compensators as they correspond to the same
BRST complex. In all the examples of specific models with unfree
gauge symmetry reviewed in Section 6, the compensator fields are
known. From the viewpoint of Section 5, it looks as an expected fact
rather than a coincidence. The reinterpretation of the anti-fields
$\xi$ to the equations constraining the ghosts (\ref{gh-constr}) as
a compensators can be further extended, in principle, in another
direction. The Lagrangian equations can have the lower order
\emph{differential} consequences. This is a typical case for the
Lagrangian theories having the second class constraints in the
Hamiltonian formalism, for example. Then, Lagrangaian equations do
not constitute the involutive PDE system. Concerning the specifics
of (non-)involutive equations, see \cite{Involution}. In particular,
the theory admits the implicit Noether identities which involve the
original Lagrangian equations and lower order consequences. These
consequences and identities control the degree of freedom number on
equal footing with the original equations and their gauge
symmetries. So, the consistent deformation of the non-involutive
Lagrangian theory is not controlled by the naive BV master equation
once it does not respect the consequences and implicit identities
\cite{Involution}. From the algebraic standpoint, the implicit
identities are similar to the modified Noether identities (\ref{GI})
if the completion functions $\tau$ are replaced by the lower order
differential consequences of the Lagrangian equations. If the BV
embedding procedure of Section 3 is aplied to the non-involutive
Lagrangian system, with $\tau_a$ being the lower order consequences
of the Lagrangian equations, the variables $\xi^a$ would play the
role of the St\"uckelberg fields. So, this BV embedding algorithm
would work as a systematic procedure of consistent inclusion of
St\"uckelberg fields. In terms of Hamiltonian formalism, the general
methods are known of conversion the second class constraints into
the first class ones, \cite{Batalin:1991jm}, \cite{Batalin:2005df},
while at Lagrangian level the ``St\"uckelbergization" is rather a
series of ad hoc tricks adjusted to specific models than a general
systematic method. We expect that the BV embedding procedure of
Section 3 can be reshaped into the general method of
``St\"uckelbergization" in the Lagrangian formalism. One more aspect
of the general connection between the theory with unfree gauge
symmetry (hence, with nontrivial completion functions) and its
equivalent with the compensator fields and without constraints on
the gauge parameters is related to the ``global modes''. As one can
see from the examples, the completion functions usually reduce to
the arbitrary constants on shell (see in Section 6). With the fixed
boundary conditions imposed on the fields, these constants take
fixed values defined by the boundary. If the setup is adopted with
unfixed boundary conditions for the fields, these constants would
play the role of the global conserved quantities. The corresponding
global degrees of freedom can be understood as modular parameters.
These degrees of freedom have to be accounted for in the BV
formalism once the fields are not fixed at the boundary. This issue
is not addressed in the present work, though the formalism can
accommodate the modular parameters, in principle.

\vspace{0.2cm} \noindent \textbf{Acknowledgements}. We thank
 A.~Sharapov for discussions. The work is partially supported by Tomsk State
University Competitiveness Improvement Program. The work of SLL is
supported by the project 3.5204.2017/6.7 of Russian Ministry of
Science and Education.

\end{document}